\documentclass[a4paper, 12pt]{article}

\usepackage{graphics}
\usepackage{floatflt, graphicx}
\usepackage{amssymb}
\usepackage{subfigure}

\begin{document}

\title {Lectures on High-Energy Neutrino Astronomy\footnotemark}
\author{Francis Halzen\\
Department of Physics, University of Wisconsin, Madison, WI 530706}

\date{}

\maketitle

\footnotetext{Lectures presented at the International WE Heraeus Summer School on ÒPhysics with Cosmic 
AcceleratorsÓ, Bad Honnef, Germany (2004)}

\begin{abstract}
Kilometer-scale neutrino detectors such as IceCube are discovery instruments covering nuclear and particle physics, cosmology and astronomy. Examples of their multidisciplinary missions include the search for the particle nature of dark matter and for additional small dimensions of space. In the end, their conceptual design is very much anchored to the observational fact that Nature produces protons and photons with energies in excess of $10^{20}$ and $10^{13}$\,eV, respectively. The cosmic ray connection sets the scale of cosmic neutrino fluxes. In this context, we discuss the first results of the completed AMANDA detector and the science reach of its extension, IceCube. Similar experiments are under construction in the Mediterranean. Neutrino astronomy is also expanding in new directions with efforts to detect air showers, acoustic and radio signals initiated by super-EeV neutrinos. The outline of these lectures is as follows:
\begin{itemize}
\item Introduction
\item Cosmic Neutrinos Associated with the Highest Energy Cosmic Rays
\item Why Kilometer-Scale Detectors?
\item Blueprints of Cosmic Accelerators: Gamma Ray Bursts and Active Galaxies
\item High Energy Neutrino Telescopes: Methodologies of Neutrino Detection
\item High Energy Neutrino Telescopes: Status
\end{itemize}
\end{abstract}

\section{Introduction}

High energy neutrino astronomy exists; we have observed the Sun and a supernova in 1987. Each observation has been deservedly rewarded with a Nobel Prize. These achievements were monumental. After thirty years the solar neutrino puzzle was resolved by the discovery that neutrinos have mass. The skeptics were proven wrong, John Bahcall knows how the Sun shines. Some 20 supernova neutrinos where adequate to confirm the basic theoretical picture of the death of a star. The goal of neutrino telescopes is to look beyond the Sun, possibly to the edge of the Universe. Construction of IceCube and other high-energy neutrino telescopes is mostly motivated by their potential to open a new window on the Universe using neutrinos as cosmic messengers. This will be the central topic of these lectures.

We should not forget the wealth of particle physics that was obtained with a small sample of supernova neutrino events, and the most exciting result of solar neutrino astronomy is not related to the Sun but to the neutrino itself. With the discovery of neutrino mass in underground experiments,
particle physics reconnected with the early cosmic ray tradition of
doing basic particle physics with heavenly beams. The original results
have since been confirmed by experiments using ``man-made'' neutrinos
from accelerators and nuclear reactors\cite{review}.  We are now
entering an era of precision neutrino physics and, one may assume, an
era of accelerator-based physics. This may not be completely true because
of the potential of IceCube to collect high statistics samples of
atmospheric neutrinos.  Especially unique is their energy range
covering $0.1 \sim 10^4$~TeV, not within reach of accelerators. Cosmic
beams of even higher energy may exist, but the atmospheric beam is
guaranteed.

Construction of IceCube and other high-energy neutrino telescopes is
mostly motivated by their potential to open a new window on the
Universe using neutrinos as cosmic messengers. The IceCube experiment
nevertheless appeared on the U.S.\ Roadmap to Particle
Physics\cite{baggerbarish}, and, as we will argue, deservedly so. As
the lightest of fermions and the most weakly interacting of particles,
neutrinos occupy a fragile corner of the Standard Model and one can
realistically hope that they will reveal the first and most dramatic
signatures of new physics.

IceCube's opportunities for particle physics are only limited by
imagination; they include
\begin{enumerate}
  \item The search for neutrinos from the annihilation of dark matter
    particles gravitationally trapped at the center of the Sun and the
    Earth.
  \item The search for theories where particle interactions, including
    gravity, unify at the TeV scale.  Neutrinos with energies
    approaching this scale will interact gravitationally with large
    cross sections, similar to those of quarks and leptons, and this
    increase should yield dramatic signatures in a neutrino telescope
    including, possibly, the production of black holes.
  \item The search for deviations from the neutrino's established
    oscillatory behavior that result from non-standard interactions,
    for instance neutrino decay or quantum decoherence.
  \item The search for a breakdown of the equivalence principle as a
    result of non-universal interactions with the gravitational field
    of neutrinos with different flavors.
  \item Similarly, the search for breakdown of Lorentz invariance
    resulting from different limiting velocities of neutrinos of
    different flavors.
  \item The search from particle emission from cosmic strings or other
    topological defects and heavy cosmological remnants created in the
    early Universe.
  \item The search for magnetic monopoles.
\end{enumerate}

As with conventional astronomy, we have to observe the Universe looking through the atmosphere. This is a curse and a blessing; the background of neutrinos produced by cosmic rays in interactions with atmospheric nuclei provides a beam for calibration of the experiments. It also presents us with an opportunity to do particle physics\cite{GHM}; see items 3--5 above. It is
well-known that oscillations are not the only possible mechanism for
atmospheric $\nu_\mu \to \nu_\tau$ flavour
transitions\cite{npreview}. These can also be generated by
nonstandard neutrino interactions 
that mix neutrino flavours.  Examples include violations of the
equivalence principle (VEP)\cite{VEP,VEP1,qVEP}, non-standard
neutrino interactions with matter\cite{NSI}, neutrino couplings to
space-time torsion fields\cite{torsion}, violations of Lorentz
invariance (VLI)\cite{VLI1,VLI2} and of CPT
symmetry\cite{VLICPT1,VLICPT2,VLICPT3}.  Although these scenarios no
longer explain the
data\cite{oldatmfitnp,NSI2,fogli1,lipari,NSI3,fogli2}, a combined
analysis of the atmospheric neutrino and K2K data can be performed to
obtain the best constraints to date on the size of such subdominant
oscillation effects\cite{ouratmnp}.

A critical feature of these scenarios is that they introduce a
departure from the characteristic energy dependence associated with
the mass-induced oscillation wavelength\cite{yasuda1,flanagan}.  In
contrast new physics predicts neutrino oscillations with wavelengths
that are constant or decrease with energy and therefore IceCube, with
energy reach in the $0.1 \sim 10^4$~TeV range for atmospheric
neutrinos, will have unmatched sensitivity. In particular note 
that with energies of $10^3$~TeV and masses of order $10^{-2}$~eV, 
the atmospheric neutrinos in IceCube reach Lorentz factors of order 
$10^{17}$. Furthermore, for most of
this energy interval conventional oscillations are
suppressed and therefore the observation of an angular distortion of
the atmospheric neutrino flux or its energy dependence will provide
signatures for the presence of new physics mixing neutrino flavors
that are not obscured by oscillations associated with their mass.

IceCube will collect a data set of order one million neutrinos over 10 years. Not surprisingly, because of the increased energy and statistics
over present experiments, sensitivity to violations of the
equivalence principle and of Lorentz invariance, for instance, will
be improved by over two orders of magnitude; see\cite{GHM}.

Although neutrino "telescopes" are designed as discovery instruments, be it for particle or astrophysics, their conceptual design is very much anchored to the observational fact that Nature produces protons and photons with energies in excess of $10^{20}$ and $10^{13}$\,eV, respectively. The cosmic ray connection sets the scale of cosmic neutrino fluxes. We discuss this next.

\section{Cosmic Neutrinos Associated with the Highest Energy Cosmic Rays}

The flux of cosmic rays is summarized in Fig.\,1a,b\cite{gaisseramsterdam}. The  energy spectrum follows a broken power law. The two power laws are separated by a feature dubbed the ``knee''; see Fig.\,1a. Circumstantial evidence exists that cosmic rays, up to EeV energy, originate in galactic supernova remnants. Any association with our Galaxy disappears however in the vicinity of a second feature in the spectrum referred to as the ``ankle''. Above the ankle, the gyroradius of a proton in the galactic magnetic field exceeds the size of the Galaxy and it is generally assumed that we are  witnessing the onset of an extragalactic component in the spectrum that extends to energies beyond 100\,EeV. Experiments indicate that the highest energy cosmic rays are predominantly protons or, possibly, nuclei. Above a threshold of 50 EeV these protons interact with cosmic microwave photons and lose their energy to pions before reaching our detectors. This is the Greissen-Zatsepin-Kuzmin (GZK) cutoff that limits the sources to our supercluster of galaxies. 

\begin{figure}[!h]
\centering\leavevmode
\includegraphics[width=6in]{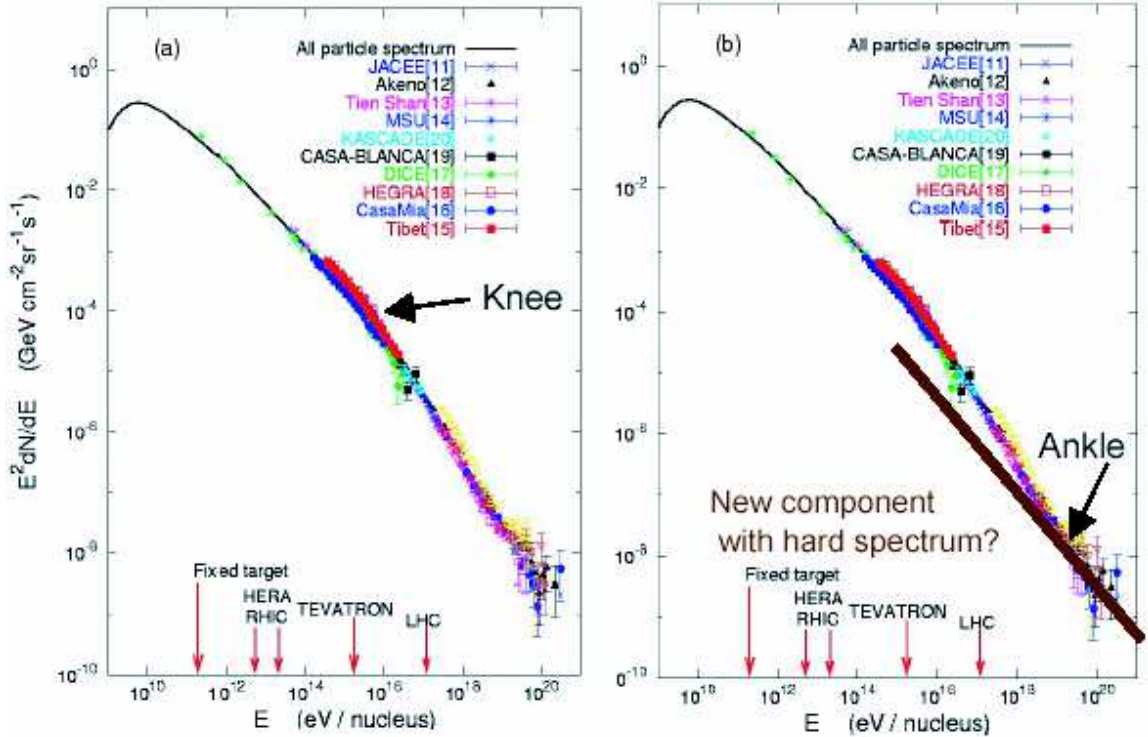}
\caption{At the energies of interest here, the cosmic ray spectrum consists of a sequence of 3 power laws. The first two are separated by the ``knee'' (left panel), the second and third by the ``ankle''. There is evidence that the cosmic rays beyond the ankle are a new population of particles produced in extragalactic sources; see right panel.}
\end{figure}

Models for the origin of the highest energy cosmic rays fall into two categories, top-down and bottom-up. In top-down models it is assumed that the cosmic rays are the decay products of cosmological remnants with Grand Unified energy scale $M_{GUT} \sim 10^{24}\rm\,eV$. These models predict neutrino fluxes most likely within reach of first-generation telescopes such as AMANDA, and certainly detectable by future kilometer-scale neutrino observatories\cite{PR}. They have not been observed.

In bottom-up scenarios it is assumed that cosmic rays originate in cosmic accelerators. Accelerating particles to TeV energy and above requires massive bulk flows of relativistic charged particles. These are likely to originate from the exceptional gravitational forces  in the vicinity of black holes. It is a fact that black holes accelerate electrons to high energy; astronomers observe them indirectly by their synchrotron radiation. We know that they accelerate protons because we detect them as cosmic rays. Because they are charged protons and deflected in magnetic fields, cosmic rays do not reveal their sources. This is the cosmic ray puzzle. Examples of candidate black holes include the dense cores of exploding stars, inflows onto supermassive black holes at the centers of active galaxies and annihilating black holes or neutron stars. Before leaving the source, accelerated particles pass through intense radiation fields or dense clouds of gas surrounding the black hole. This results in interactions producing pions decaying into secondary photons and neutrinos that accompany the primary cosmic ray beam as illustrated in Fig.\,2.

\begin{figure}[!h]
\centering\leavevmode
\includegraphics[width=4.25in]{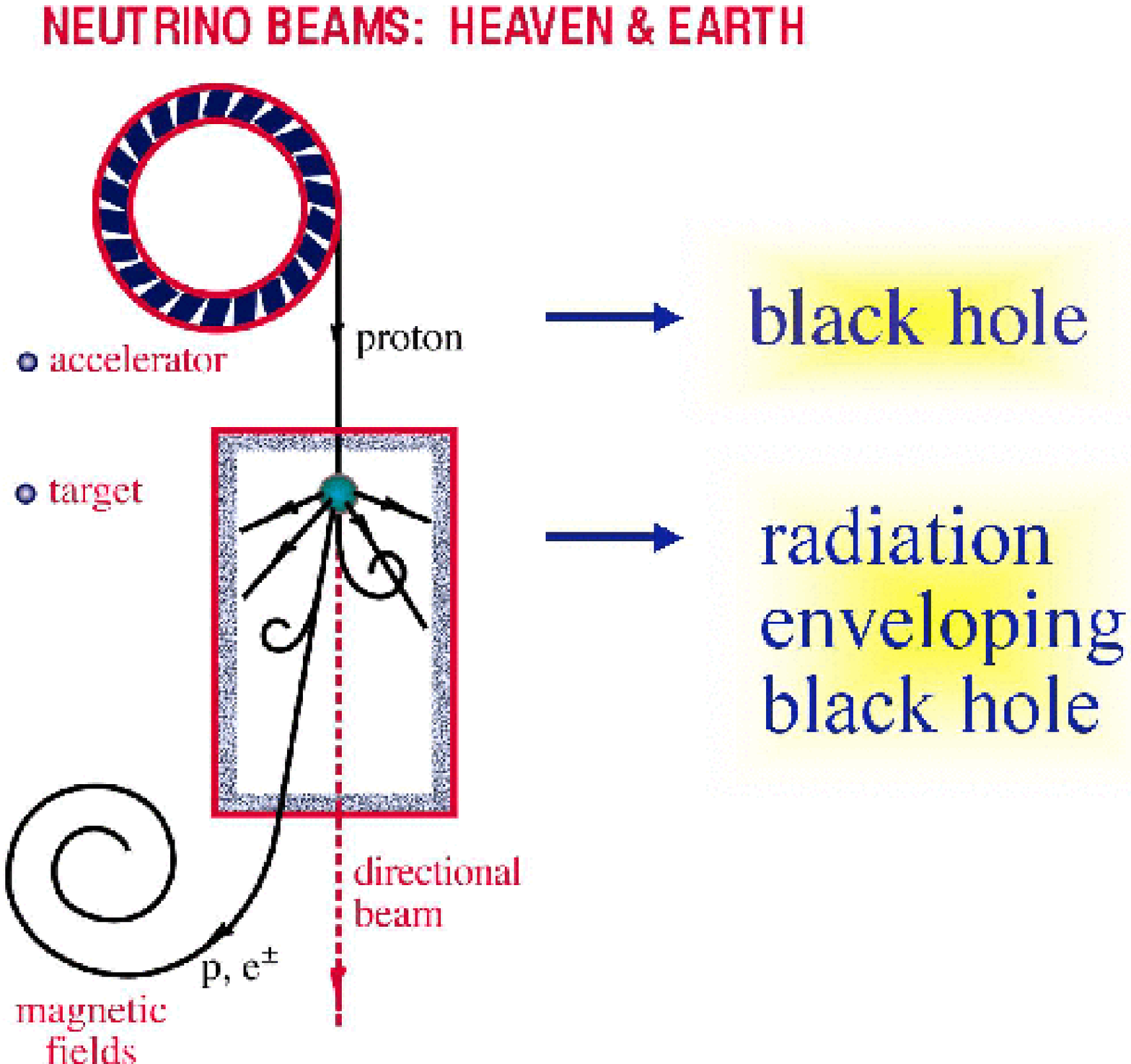}
\caption{Cosmic beam dumps exits: sketch of cosmic ray accelerator producing photons and neutrinos.}
\end{figure}

How many neutrinos are produced in association with the cosmic ray beam? The answer to this question, among many others\cite{PR},  provides the rationale for building kilometer-scale neutrino detectors. We first consider a neutrino beam produced at an accelerator laboratory; see Fig.\,2. Here the target absorbs all parent protons as well as the secondary electromagnetic and hadronic showers. Only neutrinos exit the dump. If Nature constructed such a ``hidden source'' in the heavens, conventional astronomy will not reveal it. It cannot be the source of the cosmic rays, however, because in this case the dump must be transparent to protons.

A more generic ``transparent'' source can be imagined as follows: protons are accelerated in a region of high magnetic fields where they interact with photons via the processes 
%
$p + \gamma \rightarrow \Delta \rightarrow \pi^0 + p$
and
$p + \gamma \rightarrow \Delta \rightarrow \pi^+ + n$.
%
While the secondary protons may remain trapped in the acceleration region, equal numbers of neutrons, neutral and charged pions escape. The energy escaping the source is therefore equally distributed between cosmic rays, gamma rays and neutrinos produced by the decay of neutrons and neutral and charged pions, respectively. The neutrino flux from a generic transparent cosmic ray source is often referred to as the Waxman-Bahcall flux\cite{wb1}. It is easy to calculate and the derivation is revealing.

Fig.\,1b shows a fit to the observed spectrum above the ``ankle'' that can be used to derive the total energy in extragalactic cosmic rays. The energy content of this component is derived by integrating the spectrum in Fig. 1b. It is $\sim 3 \times 10^{-19}\rm\,erg\ cm^{-3}$, assuming an $E^{-2}$ energy spectrum with a GZK cutoff. The power required for a population of sources to generate this energy density over the Hubble time of $10^{10}$\,years is $\sim 3 \times 10^{37}\rm\,erg\ s^{-1}$ per (Mpc)$^3$ or, as often quoted in the literature,
 $\sim 5\times10^{44}\rm\,TeV$ per year per (Mpc)$^3$. This works out to\cite{TKG}
\begin{itemize}
\item $\sim 3 \times 10^{39}\rm\,erg\ s^{-1}$ per galaxy,
\item $\sim 3 \times 10^{42}\rm\,erg\ s^{-1}$ per cluster of galaxies,
\item $\sim 2 \times 10^{44}\rm\,erg\ s^{-1}$ per active galaxy, or
\item $\sim 2 \times 10^{52}$\,erg per cosmological gamma ray burst.
\end{itemize}
The coincidence between these numbers and the observed output in electromagnetic energy of these sources explains why they have emerged as the leading candidates for the cosmic ray accelerators. The coincidence is consistent with the relationship between cosmic rays and photons built into the ``transparent'' source. In the photoproduction processes roughly equal energy goes into the secondary neutrons, neutral and charged pions whose energy ends up in cosmic rays, gamma rays and neutrinos, respectively.

We therefore assume that the same energy density  of $\rho_E \sim 3 \times 10^{-19}\rm\,erg\
 cm^{-3}$, observed in cosmic rays and electromagnetic energy, ends up in neutrinos with a spectrum $E_\nu dN / dE_{\nu}  \sim E^{-\gamma}\rm\, cm^{-2}\, s^{-1}\, sr^{-1}$ that continues up to a maximum energy $E_{\rm max}$. The neutrino flux follows from the relation
%
$ \int E_\nu dN / dE_{\nu}  =  c \rho_E / 4\pi  $.
%
For $\gamma = 1$ and $E_{\rm max} = 10^8$\,GeV, the generic source of the highest energy cosmic rays produces a flux of $ {E_\nu}^2 dN / dE_{\nu}  \sim 5 \times 10^{-8}\rm\, GeV \,cm^{-2}\, s^{-1}\, sr^{-1} $.

There are several ways to modify this simple prediction:
\begin{itemize} 
\item The derivation fails to take into account the fact that  there are more cosmic rays in the Universe producing neutrinos than observed at Earth because of the GZK-effect and neglects evolution of the sources with redshift. This increases the neutrino flux by a factor $\sim$\,3, possibly more.
\item For proton-$\gamma$ interactions muon neutrinos receive only 1/4 of the energy of the charged pion in the decay chain $\pi^+\rightarrow \mu^+ +\nu_{\mu}\rightarrow e^+ +\nu_e +\bar{\nu}_{\mu} +\nu_{\mu}$ assuming that the energy is equally shared between the 4 leptons and taking into account that oscillations over cosmic distances distribute the neutrino energy equally among the 3 flavors.
\end{itemize}
The corrections approximately cancel. The transition from galactic to extragalactic sources is debated; a transition at lower energy significantly increases the energy in the extragalactic component. This raises the possibility of an increase in the associated neutrino flux\cite{ringwald}.

Studying specific models of cosmic ray accelerators one finds that the energy supplied by the black hole to cosmic rays usually exceeds that transferred to pions, for instance by a factor 5 in the case of gamma ray bursts. We therefore estimate that the muon-neutrino flux associated with the sources of the highest energy cosmic rays is loosely confined to the range $ {E_\nu}^2 dN / dE_{\nu}= 1\sim 5 \times 10^{-8}\rm\, GeV \,cm^{-2}\, s^{-1}\, sr^{-1} $ yielding $10 \,{\sim}\, 50$ detected muon neutrinos per km$^2$ per year. This number depends weakly on $E_{\rm max}$ and the spectral slope~$\gamma$. The observed event rate is obtained by folding the predicted flux with the probability that the neutrino is actually detected in a high energy neutrino telescope; only one in a million is at TeV energy. The probability is given by the ratio of the muon and neutrino interaction lengths in the detector medium, $\lambda_\mu / \lambda_\nu$\cite{PR}; this will be explained in the section on methodologies.

This flux has to be compared with the sensitivity of ${\sim}10^{-7}\rm\, GeV\ cm^{-2}\, s^{-1}\,sr^{-1}$ reached during the first 4 years of operation of the completed AMANDA detector in 2000--2003\cite{hill}. The analysis of the data has not been completed, but a preliminary limit of  $2.9 \times 10^{-7}\rm\,GeV\ cm^{-2}\,s^{-1}\,sr^{-1}$ has been obtained with a single year of data\cite{b10-diffuse}. On the other hand, after three years of operation IceCube will reach a diffuse flux limit of $E_{\nu}^2 dN / dE_{\nu} = 2\,{\sim}\, 7 \times 10^{-9}\rm\,GeV \,cm^{-2}\, s^{-1}\, sr^{-1}$. The exact value depends on the magnitude of the dominant high energy atmospheric neutrino background from the prompt decay of atmospheric charmed particles\cite{ice3}. The level of this background is difficult to anticipate. A cosmic flux at the ``Waxman-Bahcall" level will result in the observation of several hundred neutrinos in IceCube\cite{ice3}.

\section{Why Kilometer-Scale Detectors?}

Arguing that a generic cosmic accelerator produces equal energies in cosmic rays, photons and neutrinos, we derived the  ``Waxman-Bahcall'' flux. A kilometer-scale detector is required to detect the roughly $10 {\sim} 50$ events per km$^2$ year. Model calculations assuming that active galaxies or gamma-ray bursts are the actual sources of cosmic rays yield similar, or even smaller event rates.

The case for kilometer-scale detectors also emerges from the consideration of  ``guaranteed'' cosmic fluxes. Neutrino fluxes are guaranteed when both the accelerator and the pion producing target material can be identified. Examples include:
\begin{itemize}
\item The extragalactic cosmic rays produce $0.1 \sim$ a few events per km$^2$ year in interactions with cosmic microwave photons\cite{cos1,cos2}. Furthermore, these cosmic rays are magnetically trapped in galaxy clusters and produce additional neutrinos on the X-ray emitting gas in the cluster.
\item Galactic cosmic rays interact with hydrogen in the disk producing an observable neutrino flux in a kilometer-scale detector.
\item Air shower arrays have observed a ``directional'' flux of cosmic rays from the galactic plane, unlikely to be protons whose directions are scrambled in the magnetic field. The flux appears only in a narrow energy range from $1\,{\sim}\, 3$\,EeV, the energy where neutrons reach typical galactic kiloparsec distances within their lifetime of minutes. Both the directionality and the characteristic energy make a compelling case for electrically neutral neutron primaries. For every neutron reaching Earth, a calculable number decays into electron antineutrinos before reaching us. Their flux should be observable in neutrino telescopes\cite{luis}: from the Cygnus region at the South Pole and from the galactic center for a Mediterranean detector.
\end{itemize}
In conclusion, observation of ``guaranteed'' sources also requires  kilometer-size neutrino detectors, preferably operated over many years.

Finally and most importantly, with recent observations\cite{hess} of the supernova remnant RX J1713.7-3946 using the H.E.S.S. atmospheric Cherenkov telescope array, gamma-ray astronomy may have pointed at a truly guaranteed source of cosmic neutrinos\cite{alvarezhalzen}.  The observations of TeV-gamma rays from the supernova remnant may have identified the first site where protons are accelerated to energies typical of the main component of the galactic cosmic rays\cite{hess}. Although the resolved image of the source (the first ever at TeV energies!) reveals TeV emission from the whole supernova remnant, it shows a clear increase of the flux in the directions of known molecular clouds. This naturally suggests the possibility that protons, shock accelerated in the supernova remnant, interact with the dense clouds to produce neutral pions that are the source of the observed increase of the TeV photon signal. Furthermore, the high statistics H.E.S.S. data for the flux are power-law behaved over a large range of energies without any signature of a cutoff characteristic of synchrotron or inverse-Compton sources. Finally, follow-up observations of the source in radio-waves and X-rays have failed to identify the population of electrons required to generate TeV photons by purely electromagnetic processes; for a detailed discussion see \cite{hiraga}. Other interpretations are not ruled out\cite{hiraga} but, fortunately, higher statistics data is forthcoming.

If future data confirms that a fraction of the TeV flux of RX J1713.7-3946 is of neutral pion origin, then the accompanying charged pions will produce a guaranteed neutrino flux of roughly 20 muon-type neutrinos per kilometer-squared per year\cite{alvarezhalzen}. From a variety of such sources we can therefore expect event rates of cosmic neutrinos of galactic origin similar to those estimated for extragalactic neutrinos in the previous section. Supernovae associated with molecular clouds are a common feature of the OB associations that exist throughout the galactic plane. They have been suspected to be the sources of the galactic cosmic rays for some time.

It is important to realize that the relation between the neutrino and gamma flux is robust\cite{alvarezhalzen}. The $\nu_\mu + \bar\nu_\mu$ neutrino flux ($dN_\nu/dE_\nu$) produced by the decay of charged pions in the source can be derived from the observed gamma ray flux by imposing energy conservation:
\begin{equation}
\int_{E_{\gamma}^{\rm min}}^{E_{\gamma}^{\rm max}}
E_\gamma {dN_\gamma\over dE_\gamma} dE_\gamma = K
\int_{E_{\nu}^{\rm min}}^{E_{\nu}^{\rm max}} E_\nu {dN_\nu\over dE_\nu} dE_\nu
\label{conservation}
\end{equation}
where ${E_{\gamma}^{\rm min}}$ ($E_{\gamma}^{\rm max}$) is the minimum (maximum) energy of the photons that have a hadronic origin. ${E_{\nu}^{\rm min}}$ and ${E_{\nu}^{\rm max}}$ are the corresponding minimum and maximum energy of the neutrinos.
The factor $K$ depends on whether the $\pi^0$'s are of $pp$ or $p\gamma$ origin. Its value can be obtained from routine particle physics. In $pp$ interactions 1/3 of the proton energy goes into each pion flavor on average. In the pion-to-muon-to-electron decay chain 2 muon-neutrinos are produced with energy $E_\pi/4$ for every photon with energy $E_\pi/2$ (on average). Therefore the energy in neutrinos matches the energy in photons and $K=1$. This flux has to be reduced by a factor 2 because of oscillations. The estimate should be considered a lower limit because the observed photon flux to which the calculation is normalized may have been reduced because of absorption in the source or in the interstellar medium. 

\section{Blueprints of Cosmic Accelerators: Gamma Ray Bursts and Active Galaxies}

We have discussed generic cosmic ray accelerators and their associated neutrino fluxes. We now turn to speculations on the nature of the black hole(s). Working through specific models of cosmic ray accelerators will confirm the level of neutrino fluxes derived above from general energetics arguments. The list of speculations for the sources is long and includes, but is not limited to:
\begin{itemize}

\item{Sources Associated with Stellar Objects}

Gamma ray bursts (GRB), outshining the entire Universe for the duration of the burst, are perhaps the best motivated sources of high-energy neutrinos\cite{waxmanbahcall, mostlum2, mostlum3}; we will discuss these in detail further on. Other theorized neutrino sources associated with compact objects include supernova remnants exploding into the interstellar medium\cite{olinto,snr2,snr3,PR}, X-ray binaries\cite{PR,xray1,xray2,xray3}, soft gamma ray repeaters\cite{HLM}, mini-quasars\cite{PR,micro1,micro2}, any of which could produce observable fluxes of high-energy neutrinos.

\item{Active Galactic Nuclei (AGN)}

Blazars, the brightest objects in the Universe and the sources of TeV-energy  gamma rays, have been extensively studied as potential neutrino sources.  Blazar flares with durations ranging from months to less than an hour, are believed to be produced by relativistic jets projected from an extremely massive accreting black hole. Blazars may be the sources of the highest energy cosmic rays and, in association, produce observable fluxes of neutrinos from TeV to EeV energies.

\item{Neutrinos Associated with the Propagation of Cosmic Rays}

As previously mentioned, cosmic rays interact with the Earth's atmosphere\cite{prompt1,prompt2}, with the Sun\cite{sun1,sun2} and with the hydrogen concentrated in the galactic plane\cite{PR,plane1,plane2,plane3} producing high-energy neutrinos. Atmospheric neutrinos produce a natural neutrino beam that can be exploited for particle physics. Thus non-accelerator experiments discovered neutrino mass. Future neutrino telescopes will yield high statistics samples of neutrinos of TeV energy and above. Several million well-reconstructed neutrinos will be collected over the lifetime of the instruments.

We previously mentioned the so-called GZK neutrinos produced when extragalactic cosmic ray propagate in the microwave background. It has been speculated that cosmic neutrinos themselves may produce cosmic rays and neutrinos in interactions with relic neutrinos $\nu + \nu_b\rightarrow Z$. This is called the Z-burst mechanism \cite{z1,z2,z3}.

\item{Dark Matter, Primordial Black Holes, Topological Defects and
Top-Down Models}

The vast majority of matter in the Universe is dark with its particle nature not yet revealed.  The lightest supersymmetric
particle, or any other Weakly Interacting Massive Particles (WIMPs) proposed as particle candidates for cold dark matter, are gravitationally trapped in the Sun, Earth and the galactic center. There, they accumulate and subsequently annihilate generating high-energy neutrinos observable in neutrino telescopes\cite{ind1,ind2,ind3,ind4,ind5,ind6,ind7}. Another class of dark matter candidates are superheavy cosmic remnants with GUT-scale masses.  They may be the dark matter and their decay or annihilation products may be the source of the ultra high-energy cosmic rays. Neutrinos are also produced and the predicted fluxes are large\cite{sh1,sh2,sh3,drees,sh4}.  Extremely high-energy neutrinos are also predicted in a wide variety of other top-down scenarios invoked to explain the origin of the highest energy cosmic rays, including decaying monopoles, vibrating cosmic strings \cite{top1,top2} and Hawking radiation from primordial black holes\cite{bh1,bh2,pbh}.
\end{itemize}

In this next section we concentrate on neutrino fluxes associated with the highest energy cosmic rays. We will work through two much-researched examples: GRB and AGN. The myriad of other speculations have been recently reviewed by Learned and Mannheim \cite{PR}. We concentrate here on neutrino sources associated with the highest energy gamma rays.

\subsection{Gamma Ray Bursts}

\subsubsection{GRB Observations}

Although we do not yet understand in detail the internal mechanisms that generate GRB, the relativistic fireball model provides us with a successful phenomenology to accommodate the spectacular observations. It can be used to estimate the flux of neutrinos in a relatively model-independent fashion. The properties of GRB relevant to the calculation are:

\begin{itemize}

\item GRB are extremely luminous events releasing an energy of order of one solar mass over times that range from less than a second to tens of seconds. Evidence is accumulating that GRB are beamed and may actually be standard candles of energy $\sim 10^{52}$\,erg/s with the range of energies observed reflecting the orientation of the beam relative to the observer.

\item The durations of GRB follow a bimodal distribution with peaks near two seconds and 20 seconds, although some GRB have durations as short as milliseconds and as long as 1000 seconds\cite{dur1}.

\item Variations in the spectra occur on the scale of milliseconds\cite{msec,dur1} indicating a compact source.

\item GRB are cosmological events. Redshifts exceeding z=4 have been measured \cite{red1, red2}.

\item GRB are rare. Assuming no beaming, BATSE observed one thousand GRB per year or one burst per galaxy per million years. This rate should be increased by the beaming factor.

\item GRB produce a broken power-law spectrum of gamma rays with $\phi_\gamma\propto E_\gamma^{-2}$ for $E_\gamma \gtrsim$0.1-1 MeV and $\phi_\gamma\propto E_\gamma^{-1}$ for $E_\gamma\lesssim$ 0.1-1 MeV\cite{piran}.

\end{itemize}

The millisecond variations of the GRB flux require an original event where a large amount of energy is released in a very compact volume ($R_0\sim100\,$km). Evidence is accumulating that at least some of the GRB are associated with the collapse of massive stars to a black hole. The GRB rate is consistent with the rate of supernovae with progenitor masses exceeding several solar masses. Other speculations include the annihilation of binaries made of collapsed objects.

\subsubsection{Fireball Phenomenology}

The phenomenology that successfully describes above observations is that of a fireball expanding with highly relativistic velocity, powered by radiation pressure. The observer detects boosted energies emitted over contracted times; without this a description of the extreme observations is impossible. The dynamics of a gamma ray burst fireball is actually reminiscent of the physics of the early expanding Universe. Initially, there is a radiation dominated soup of leptons and photons and few baryons. It is hot enough to freely produce electron-positron pairs. From the observed luminosity of a burst one can derive the number density of photons $n_\gamma$:
\begin{eqnarray}
L = 4 \pi R_0^2 c n_\gamma E_\gamma,
\end{eqnarray}
where $R_0$ is the initial radius R of the source, i.e. prior to expansion.  The optical depth of a photon in the initial fireball is
determined by the photon density and the interaction cross section\cite{waxman1}:
\begin{eqnarray}
\tau_{\rm{opt}} = \frac{R_0}{\lambda_{\rm{int}}} = R_0 n_\gamma
\sigma_{\rm{Th}}
=\frac{L\sigma_{\rm{Th}}}{4\pi R_0
c E_\gamma}\sim 10^{15}\bigg(\frac{L_{\gamma}}{10^{52}
\rm{erg/s}}\bigg)\bigg(\frac{100 \,\rm{km}}{R_0}\bigg)\bigg(\frac{1\,
\rm{MeV}}{E_\gamma}\bigg).
\end{eqnarray}
Here $\lambda_{\rm{int}}$ is the interaction length of a photon as a result of pair production and Thomson scattering. These cross sections are roughly equal, with the Thomson cross section $\sigma_{\rm{Th }}\simeq 10^{-24}\rm{cm}^2$.

With an optical depth of order $\sim 10^{15}$, photons are trapped in the fireball. It cannot radiate. This causes the highly relativistic expansion of the fireball powered by radiation pressure\cite{fire1,fire2,fire3,pro7}. The fireball will expand with increasing velocity until it becomes transparent and the radiation is released in the visual display of the GRB. By this time, the expansion velocity has reached highly relativistic values of order $\gamma\simeq300$ as we will deduce further on. \footnote{For a specific (beamed) GRB the boost factor $\gamma$ is related to the Doppler factor which depends on its redshift z and orientation $\theta$ relative to the observer: D $= [(1+z)(1- \beta cos\theta)\gamma]^{-1}$ with $\gamma = (1-\beta^2)^{-\frac{1}{2}}$.}

Besides leptons and photons, the fireball contains some baryons. During expansion, the opaque fireball cannot radiate and any nucleons present are accelerated as radiation is converted into bulk kinetic energy. It can be shown that, with expansion, the $\gamma$-factor grows linearly with R until reaching the maximum value $\eta$ when the radiation is emitted. With the release of the radiation, there is a transition from radiation to matter dominance of the fireball. At this stage, the radiation pressure is no longer important and the expanding fireball coasts without acceleration. The expansion velocity remains constant with $\gamma\simeq \eta \equiv\frac{L}{\dot{M}c^2}$ that is determined by the amount of baryonic matter present, often referred to as the baryon loading\cite{fire2,fire3}.
\begin{figure}[!h]
\centering\leavevmode
\includegraphics[width=5in]{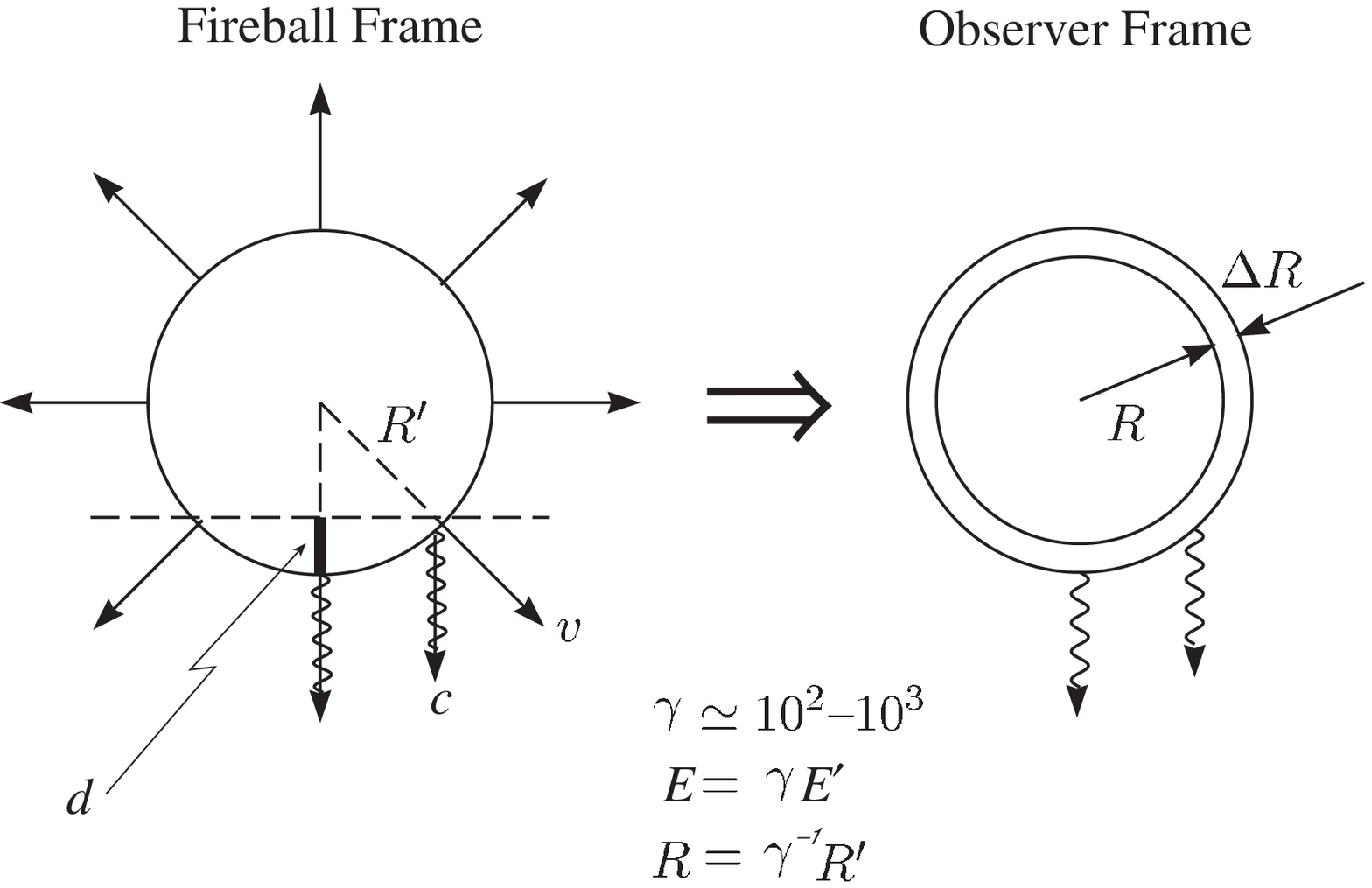}
\caption{Diagram of GRB fireball kinematics assuming no beaming. Primed
quantities refer to the comoving frame. Unprimed quantities refer to the
observer's frame.}
\label{seventeenb}
\end{figure}
The phenomenology will reveal values of $\eta$ between $10^2$ and $10^3$\cite{waxman1,waxman2,piran}. The formidable appearance of the GRB display is simply associated with the large boost between the fireball and the observer who detects highly boosted energies and contracted times.

The exploding fireball's original size, $R_0$, is that of the compact progenitor, for instance the black hole created by the collapse of a massive star. As the fireball expands the flow is shocked in ways familiar from the emission of jets by the black
holes at the centers of active galaxies or by solar mass black holes in mini-quasars. (A possible way to visualize the formation the latter is to imagine that infalling material accumulates and chokes the black hole. At this point a blob of plasma is ejected. Between these ejections the emission is reduced.) The net result is that the expanding fireball is made up of multiple shocks. These are the sites of the acceleration of particles to high-energy and the seeds for the complex millisecond structures observed in individual bursts. When the shocks expand with a range of velocities, they collide providing a mechanism to accelerate particles to high-energy. The characteristic width of these shocks in the fireball frame is $\delta R^\prime=\gamma c \Delta t$, where $\Delta t \simeq 0.01\,$sec reflecting the very compact progenitor.

The expanding shock is seen by the observer as an expanding shell of thickness $c\Delta t=R_0$ and radius $R$; see Fig.\,3. $\Delta t$ the time scale of fluctuations in the fireball; it is related to $R^\prime$ by:
\begin{eqnarray}
R^\prime = \gamma^2 c \Delta t = \gamma^2 R_0,
\end{eqnarray}
with primed quantities referring from now on to the frame where the fireball is at rest. Two, rather than a single $\gamma$-factor, relate the two quantities because of the geometry that relates the radius $R^\prime$ to the time difference between  photons emitted from a shell expanding with a velocity $v$; see Fig.\,3\cite{halzenlec}. Introducing the separation $d$ of the two photons along the line of sight, we note that
\begin{eqnarray}
\Delta t \sim\frac{d}{c}\sim\frac{1}{c}(R^\prime-
R^\prime\frac{v}{c})\simeq
\frac{R^\prime}{2c}(1-\frac{v^2}{c^2})\sim\frac{R^\prime}{c\gamma^2},
\end{eqnarray}
using the relativistic approximation that $1+\frac{v^2}{c^2}\simeq1$.

\subsubsection{The Observed GRB Spectrum: Synchroton and, Possibly,  Inverse Compton Photons and Cosmic Rays}

The observed broken power-law gamma ray spectrum of GRB, with two distinct spectral slopes, is far from a blackbody spectrum indicating that fireball photons do not sufficiently interact to thermalize prior to escaping the fireball. After escaping, the photons show spectral features characteristic of the high-energy, non-thermal emission by supernova remnants and active galaxies. Here photons up to MeV-energy can be produced by synchrotron radiation, with some reaching, possibly, up to TeV energies by inverse Compton scattering on accelerated electrons\cite{piran,milagro}. These processes have
been modeled for the expanding fireball and successfully accommodate observed GRB spectra. To produce the non-thermal spectrum, special conditions must prevail\cite{waxman1,waxman2,halzenlec,piran}. The photons must not thermalize before the shock becomes transparent so that the observed radiation is indeed non-thermal. Conversely, if the GRB is released too early, there is insufficient time for the synchotron and inverse Compton scattering processes to produce the observed spectrum. This delicate balance fixes the $\gamma$ factor in a narrow range of values in the interval $10^2-10^3$. The derivation is straightforward; see reference\cite{waxman2,piran}. This represents a major success of the relativistic fireball phenomenology\cite{syn1,syn2,syn3,syn4,syn5}.

It may be possible that protons are accelerated to energies above $10^{20}$\,eV in the fireball that accommodates the GRB observations\cite{waxman2,waxman2b,waxman2c}. GRB within a radius of 50-100 Mpc over which protons can propagate in the microwave background, may therefore be the sources of the ultra high-energy cosmic rays (UHECR's)\cite{dermer, waxman2,waxman2b,waxman2c,
waxman3, uhecr1,uhecr2,uhecr3,mostlum3,olinto}.  To accelerate protons to this energy, two conditions have to be satisfied:

\begin{itemize}

\item The acceleration time $t_a \sim R_L/c$, where $R_L=E/eB\gamma$ is the Larmor radius, must not exceed the duration of the burst $R/\gamma c$,
\begin{eqnarray}
\frac{E}{\gamma e B} \lesssim \frac{R}{\gamma},
\end{eqnarray}
\item Energy losses due to synchrotron radiation must not exceed the energy gained by acceleration.
\begin{eqnarray}
t_{syn}=\frac{6 \pi }{\sigma_T m_e^2 c^5 B^2 \gamma_p}=\frac{6 \pi
m_p^4c^3}{\sigma_T m_e^2 E B^2},         
\end{eqnarray}
where $\gamma_p$ is related to the photon energy $\gamma_p = \frac{E}{m_p}$.

\end{itemize}

Combining above requirements
\begin{eqnarray}
\bigg(\frac{E}{10^{20}\rm{eV}}\bigg)
\bigg(\frac{10^{11}\rm{m}}{R}\bigg) \rm{ tesla} \lesssim
\bigg(\frac{B}{10\rm{T}}\bigg) \lesssim \bigg(\frac{\gamma}{300}\bigg)^2
\bigg(\frac{10^{20}\rm{eV}}{E}\bigg)^2,
\end{eqnarray}
or
\begin{eqnarray}
R \gtrsim 10^{10} \, \rm{ meters} \bigg(\frac{300}{\gamma}\bigg)^2
\bigg(\frac{E}{10^{20}\rm{eV}}\bigg)^3.
\end{eqnarray}
From simple fireball kinematics, we previously derived that
\begin{eqnarray}
R \lesssim \gamma^2 c \Delta t,
\end{eqnarray}
where $\Delta t$ is $\sim10$ msec. Combining this with Eq.\,34 leads to the final requirement:
\begin{eqnarray}
10^{10} \bigg(\frac{300}{\gamma}\bigg)^2
\bigg(\frac{E}{10^{20}\rm{eV}}\bigg)^3 \lesssim R \lesssim \gamma^2 c
\Delta t,
\end{eqnarray}
or
\begin{eqnarray}
\gamma \gtrsim 130 \bigg(\frac{E}{10^{20}\rm{eV}}\bigg)^{3/4}
\bigg(\frac{.01 \rm{sec}}{\Delta t}\bigg)^{1/4},
\end{eqnarray}
which can indeed be satisfied with the same range of values of $\gamma=10^2-10^3$ required to accommodate the observed photon spectrum.

We conclude that bursts with Lorentz factors $\gtrsim 300$ can accelerate protons to $\sim 10^{20}$ eV. The long acceleration time of 10-100 seconds implies however that the fireball extends to a large radius where interstellar matter may play an important role in the kinematics of the expanding shell.

We showed that the shocked protons do not loose their energy by synchrotron radiation. One can verify that energy losses from $p-\gamma$ interactions will also not interfere with acceleration to high-energy. These will, in fact, be a source of high-energy neutrinos associated with the beam of high-energy protons. We will discuss this next.

\subsubsection{Neutrino Production in GRB Fireballs}

While the fireball energy transferred to highly relativistic electrons is observed in the form of radiation, it is a matter of speculation how much energy is transferred to protons. The assumption that GRB are the sources of the highest energy cosmic rays does determine the energy of the fireball baryons. Accommodate the observed cosmic ray spectrum of extragalactic cosmic rays requires roughly equal efficiency for conversion of fireball energy into the kinetic energy of protons and electrons.

In this scenario the production of PeV neutrinos in the GRB fireball is a robust prediction\cite{waxmanbahcall, halzenlec, dermer}. Neutrinos are inevitably produced in interactions of accelerated protons with fireball photons, predominantly via the processes
\begin{eqnarray}
p\gamma \rightarrow \Delta \rightarrow n \pi^{+}
\end{eqnarray}
and
\begin{eqnarray}
p\gamma \rightarrow \Delta \rightarrow p \pi^{0}.
\end{eqnarray}
These have large cross sections of order $10^{-28} \rm{cm}^2$.  The charged $\pi$'s subsequently decay producing charged leptons and neutrinos, while neutral $\pi$'s generate high-energy photons that may be observable in TeV energy air Cerenkov detectors. For the center-of-mass energy of a proton-photon interaction to exceed the threshold energy for producing the $\Delta$-resonance, the comoving proton energy must exceed
\begin{eqnarray}
E^\prime_p > \frac{m_\Delta^2-m_p^2}{4 E^\prime_\gamma}.
\end{eqnarray}
Therefore, in the observer's frame,
\begin{eqnarray}
E_p > 1.4 \times 10^{16} \rm{eV} \bigg(\frac{\gamma}{300}\bigg)^2
\bigg(\frac{1 MeV}{E_\gamma}\bigg),
\end{eqnarray}
resulting in a neutrino energy
\begin{eqnarray}
E_\nu = \frac{1}{4} \langle x_{p \rightarrow \pi} \rangle E_p > 7 \times
10^{14} \rm{eV} \bigg(\frac{\gamma}{300}\bigg)^2 \bigg(\frac{1
MeV}{E_\gamma}\bigg).
\end{eqnarray}
Here $\langle x_{p \rightarrow \pi} \rangle \simeq .2$ is the average fraction of energy transferred from the initial proton to the produced pion.  The factor of 1/4 is based on the estimate that the 4 final state leptons in the decay chain $\pi^{\pm} \rightarrow\bar{\nu_{\mu}} \mu \rightarrow\bar{\nu_{\mu}} e \nu_e \bar{\nu_e}$ equally share the pion energy.

As the kinetic energy in fireball protons increases with expansion, a fraction of their energy is converted into pions once the protons are accelerated above threshold for pion production. The fraction of energy converted to pions $f_\pi$ depends on the number of proton mean-free paths within a fireball of radius $\Delta R^\prime$. Referring to the fireball kinematics previously introduced, the width of a shock in the fireball frame is $\Delta R^\prime=\gamma c \Delta t$, where $\Delta t \simeq 0.01\,$sec. We
obtain:
\begin{eqnarray}
f_{\pi} \simeq \frac{\Delta R^\prime}{\lambda_{p \gamma}}\langle
x_{p\rightarrow \pi}\rangle.
\end{eqnarray}
Here the proton interaction length $\lambda_{p \gamma}$ in the photon fireball is given by
\begin{eqnarray}
\frac{1}{\lambda_{p\gamma}}=n_\gamma \sigma_{\Delta}.
\end{eqnarray}
$n_\gamma$ is the number density of photons in the fireball frame and $\sigma_\Delta \sim 10^{-28} \rm{cm}^2$ is the proton-photon cross section at the $\Delta$-resonance.  The photon number density is the ratio of the photon energy density and the average photon energy in the comoving frame:
\begin{eqnarray}
n_\gamma=\frac{U_\gamma^\prime}{E_{\gamma}^\prime}=\bigg(\frac{L_\gamma
\Delta t/\gamma}{4\pi R^{\prime 2} \Delta R^\prime}\bigg)   \bigg/
\bigg(\frac{E_\gamma}{\gamma}\bigg).
\end{eqnarray}
Using the fireball kinematics of Eqs.\,15 and 16
\begin{eqnarray}
n_\gamma=\bigg(\frac{L_\gamma}{4\pi c^3 \Delta t^2
\gamma^6}\bigg)   \bigg/
\bigg(\frac{E_\gamma}{\gamma}\bigg)=\frac{L_\gamma}{4\pi c^3 \Delta t^2
\gamma^5 E_{\gamma}}.
\end{eqnarray}
Note that for $\gamma=1$ this equation recovers Eq. 2 with $R_{0} = c \Delta t$; this represents the essence of the fireball phenomenology, namely the requirement of highly relativistic boost factors for achieving transparency to radiation. It also determines the fraction of proton energy converted to $\pi$'s in the expansion:
\begin{eqnarray}
f_{\pi} \simeq \frac{L_{\gamma}}{E_{\gamma}}\frac{1}{\gamma^4 \Delta t}
\frac{\sigma_\Delta \langle x_{p \rightarrow \pi} \rangle}{4 \pi c^2}
\simeq .13 \times
\bigg(\frac{L_\gamma}{10^{52}\rm{erg/s}}\bigg) \bigg(\frac{1
\rm{MeV}}{E_\gamma}\bigg)  \bigg(\frac{300}{\gamma}\bigg)^4
\bigg(\frac{.01 \rm{sec}}{\Delta{t}}\bigg).
\end{eqnarray}
We therefore conclude that for $L \sim 10^{52}$ erg/sec, $\Delta t \sim 10$ msec and $\gamma \simeq$ 300, the fraction of fireball energy converted to pions is on the order of 10 percent. This quantity strongly depends on the Lorentz factor $\gamma$.

In order to normalize the neutrino flux we introduce the assumption that GRB are the source of cosmic rays above the ankle of the cosmic ray spectrum near $\sim 3 \times 10^{18}$\,eV \cite{dermer, waxman2,waxman2b,waxman2c, waxman3,uhecr1,uhecr2,uhecr3,mostlum3}.  The flux in neutrinos can then be simply obtained from the total energy injected into cosmic rays and the average energy of a single neutrino\cite{waxmanbahcall}:
\begin{eqnarray}
\phi_\nu \simeq
\frac{c}{4\pi}\frac{U^\prime_\nu}{E^\prime_\nu}=\frac{c}{4
\pi}\frac{U_\nu}{E_\nu}=\frac{c}{4 \pi}\frac{1}{E_\nu}\bigg( \frac{1}{2}
f_\pi t_H \frac{dE}{dt} \bigg),
\end{eqnarray}
or
\begin{eqnarray}
\phi_\nu = 2 \times 10^{-14} \rm{cm}^{-2} \rm{s}^{-1} \rm{sr}^{-1} \bigg(
\frac{7 \times 10^{14}\rm{eV}}{E_\nu} \bigg) \bigg(\frac{f_\pi}{0.125}
\bigg) \bigg(\frac{t_H}{10 \rm{Gyr}} \bigg) \bigg( \frac{dE/dt}{4 \times
10^{44}\, \rm{Mpc}^{-3} \rm{yr}^{-1}} \bigg),
\end{eqnarray}
where $t_H \sim 10$ Gyrs is the Hubble time and $dE/dt \sim 4 \times 10^{44}\, \rm{ergs}\,\rm{Mpc}^{-3}\,\rm{yr}^{-1}$ is the observed injection rate of energy into the Universe in the form of cosmic rays above the ankle. This injection rate was already introduced in section 2.

PeV neutrinos are detected by observing a charged lepton produced in the charged current interaction of a neutrino near the detector. For instance, the probability $P_{\nu \rightarrow \mu}$  to detect a muon neutrino within a neutrino telescope's effective area is given by $\lambda_\mu  / \lambda_\nu$ as previously discussed. At TeV-PeV energies the function $ P_{\nu \rightarrow \mu}$ can be approximated by
\begin{eqnarray}
P_{\nu \rightarrow \mu} \simeq  1.7 \times 10^{-6}
E^{0.8}_{\nu,obs}(\rm{TeV}),
\end{eqnarray}
where $E_{\nu,obs}=E_{\nu}/(1+z)$ is the observed neutrino energy.  The rate of detected events is the convolution of the flux with the probability of detecting the neutrino (details follow in section 5)
\begin{eqnarray}
N_{events} = \int_{E_{\rm{thresh}}}^{E_\nu^{\rm{max}}} \phi_\nu P_{\nu
\rightarrow \mu} \frac{dE_\nu}{E_\nu} \simeq 25 \,\rm{km}^
{-2}\rm{yr}^{-1}
\end{eqnarray}
With the ability to look for GRB neutrino events in coincidence with gamma ray observations, i.e. in short time windows over which very little background accumulates in the neutrino detector, there is effectively no background for this neutrino signature of GRB.

Above energetics argument yields the diffuse flux of neutrinos produced collectively by all GRB. Although kilometer-scale neutrino telescopes have the capability to observe this flux, it is advantageous to make coincident observations with satellite experiments in order to reduce background. Such measurements are sensitive to the burst-to-burst fluctuations of the neutrino flux resulting from variations in distance and energy of individual GRB. It can be shown that, in the presence of fluctuations, the ``coincident" flux is produced by a few special bursts\cite{fluc1,fluc2}, the ``average" burst does not contribute. Above derivation of the neutrino flux additionally neglects the redshift evolution of the sources that is believed to approximately follow that of starburst galaxies. These considerations have the tendency to reduce the prediction derived above, but not by much\cite{guetta,becker}.

\subsubsection{Other Opportunities for Neutrino Production in GRB}

There are several other opportunities for neutrino production following the chain of events in the lifetime of the fireball.

\begin{itemize}

\item The core collapse of massive stars has emerged as the likely origin of the ``long" GRB with durations of tens of seconds. The fireball produced is likely to be beamed in jets along the rotation axis of the collapsed object. The mechanism is familiar from observations of jets associated with the central black hole in active galaxies. The jets eventually run into the stellar material that is still accreting onto the black hole. If the jets successfully puncture through this stellar envelope they will emerge to produce a GRB. While the fireball penetrates the remnant of the star, the fast particles in the tail will catch up with the slow particles in the leading edge and collide providing an opportunity for pion production yielding neutrinos of tens of TeV energy\cite{waxmantev1,waxmantev2}.  Interestingly, failed jets that do not emerge will not produce a visible GRB but do produce observable neutrinos. Bursts within a few hundred megaparsecs ($\sim 10$ bursts per year as well as an additional unknown number of ``invisible'' bursts from failed GRB) may actually produce large rates of TeV neutrino events in a kilometer-scale detector and, possibly, observable rates in a first-generation detector such as AMANDA\cite{beacom}.

\item Afterglow observations show that external shocks are produced when the GRB runs into the interstellar medium, including a reverse shock that propagates back into the burst ejecta.  Electrons and positrons in the reverse shock radiate an afterglow of eV-keV photons that represent a target for neutrino production by ultra high-energy protons accelerated in the burst. One expects EeV neutrinos with rather low rates, of order $0.01\rm \ km^{-2} {yr}^{-1}$. This mechanism may represent an opportunity for detectors with a high threshold, but also larger effective area, to do GRB physics. The neutrino energy may indeed be above the threshold for EeV telescopes using acoustic, radio or horizontal air shower detection
techniques. 

\item Finally, the conversion of radiation into kinetic energy in the fireball will accelerate neutrons along with protons, especially if
the progenitor involves neutron stars.  Protons and neutrons are initially coupled by nuclear elastic scattering. If the expansion of the fireball is sufficiently rapid the neutrons and protons will no longer interact. Neutrons decouple from the fireball while protons are still accelerated. Protons and neutrons may thus achieve relative velocities sufficient to generate pions which decay into GeV neutrinos \cite{gev1,gev2}.

\end{itemize}

\subsection{Blazars: the Sources of the Highest Energy Gamma rays}

\subsubsection{Blazar Characteristics}

Active Galactic Nuclei (AGN) are of special interest because some emit most of their luminosity at GeV energy and above and are therefore obvious candidate cosmic ray accelerators. A subset, called blazers, emit high-energy radiation in collimated jets pointing at the Earth and are the sources of photons with energies of tens of TeV. They have the following characteristics:

\begin{itemize}

\item
They are cosmological sources with inferred isotropic luminosities as high as $\sim10^{45}-10^{49}\rm{erg}\/\rm{s}$. They produce a typical $\phi_\gamma \propto E^{-2.2}$ spectrum in the MeV-GeV range. The energetics require a black hole one billion times more massive than our Sun.

\item
The emission displays variability over several time scales. TeV-energy bursts as short as minutes have been observed. This range of time scales indicates a sub-parsec engine; $c \Delta t \sim R^{\prime} \sim 10^{-3}-10^{-1}$ pc\cite{redagn1,histagn2,histagn22, sikora}.

\end{itemize}

\subsubsection{Blazar Models}

Blazars are powered by accreting supermassive black holes with masses of $\sim 10^7 M_{\odot}$ or larger. Some of the infalling matter is reemitted and accelerated in highly beamed jets aligned with the rotation axis of the black hole.

It is generally agreed upon that synchrotron radiation by accelerated electrons is the source of the observed IR to X-ray
flux\cite{lep1,lep2,lep3,lep4,sikora}. Inverse Compton scattering of synchrotron or, possibly, other ambient photons by the same electrons can accommodate observations in the MeV-GeV part of the spectrum. There is a competing explanation for the high energy emission\cite{had1, had2,had3,had4,had5,had6,had7,sikora}, invoking the interactions of protons, accelerated along with the electrons, with gas or radiation surrounding the black hole. Pions produced in these interactions decay into the observed gamma rays and not-yet observed neutrinos. This process is accompanied by synchrotron radiation of the protons.

\begin{figure}[!h]
\centering\leavevmode
\includegraphics[width=5in]{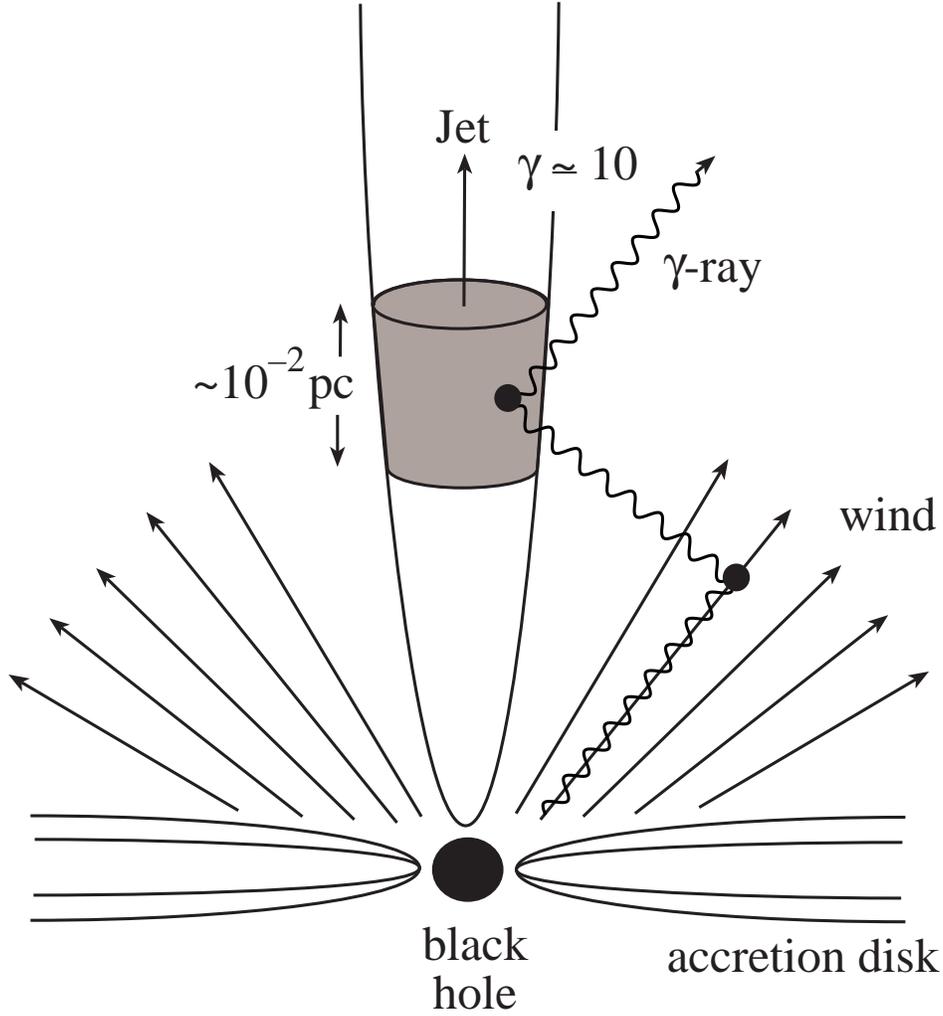}
\caption{Diagram of Blazar Kinematics}
\label{eighteen}
\end{figure}

The basic dynamics of the blazar is common to all models; see Fig. 4. Relativistic jets are generated with substructure that takes the
form of ``blobs'' or rather ``sheets'' of matter traveling along the jet with Lorentz factors that are assumed, in some cases, to reach values as large as 100. As previously discussed in the case of GRB, in order to accommodate the observation of flares, the thickness of these sheets must be less than $\gamma c \Delta t \sim 10^{-2}$ parsec, much smaller than their width, which is of the order of 1 parsec. It is in these blobs that shocks produce TeV gamma rays and high-energy neutrinos. The mechanism is, in fact, analogous to GRB although the scales are very different; see Table\,1.

\begin{table}[h]
\caption{Comparison of AGN and GRB jets}
\tabcolsep1.25em
\def\arraystretch{1.5}
\begin{center}
\begin{tabular}{|c|c|c|}
\hline
& \bf AGN& \bf GRB\\
\hline
\bf total energy&$ > 10^{45}$ erg& $10^{52}$ erg\\
\hline
\bf burst duration& minutes-days&10 msec and above\\
\hline
\bf boost factor&  < ~10& > 300\\
\hline
\bf photon energies& UV& MeV\\
\hline
\bf modelling& moving blob& expanding shell\\
\hline
\end{tabular}
\end{center}
\end{table}

Several calculations of the neutrino flux from active galactic nuclei have been performed\cite{agnnu1,agnnu2,agnnu3,agnnu4,agnnu5,agnnu6,halzenlec,berez10,dermer1, dermer2, external, external2, pohl}.  In the following sections, we describe two examples that illustrate the mechanisms for neutrino production in hadronic blazars.

\subsubsection{Shocked Protons: EeV Blazar Neutrinos}

If protons are present in blazar jets, they will interact with photons and produce neutrinos.  The process is similar to the one described for GRB shells althought the Lorentz factors are smaller and the geometry is different,  blobs moving with superluminal velocities along the jets versus expanding shells. The Lorentz factors are constrained by the energetics of the highest energy gamma rays escaping the blob. Their energy must be below threshold for $e^+e^-$ pair production with ambient photons whose density peaks in the vicinity of $\sim 10\,$ eV; the UV bump in AGN spectra. Contemplating the observations of gamma rays above 15 TeV in Markarian 501\cite{prot, highest}, evading pair production (in the blob frame) requires
\begin{eqnarray}
E_{\gamma \rm{, max}} E_{\gamma \rm{, peak}} < \gamma^2 m_e^2,
\end{eqnarray}
\begin{eqnarray}
\gamma > 25  \bigg(\frac{E_{\gamma \rm{, max}}}{15\, \rm{TeV}}\bigg)^{1/2}
\bigg(\frac{E_{\gamma \rm{, peak}}}{10\, \rm{eV}}\bigg)^{1/2}.
\end{eqnarray}
We will therefore consider Lorentz factors in the range of 10 to 100.

A main difference between blazars and GRB is the geometry of the shocks. In AGN, blobs moving along the jet axis are treated as 3D spherical balls of shocked material. This is different from the 2D expanding shells centered on the precursor of a GRB.  The energy density in a blob is
\begin{eqnarray}
U^\prime_\gamma = \frac{L^\prime_\gamma \Delta t }{\frac{4}{3}\pi
R^{\prime 3} } = \frac{L_\gamma \Delta t }{\gamma \frac{4}{3}\pi
(\gamma c
\Delta t)^3 } = \frac{3 L_\gamma}{4 \pi c^3 \gamma^4 \Delta t^2}
\end{eqnarray}
Except for geometry, this result is identical to Eq.\,20. We obtain a number density of photons
\begin{eqnarray}
n_\gamma =\frac{U^\prime_\gamma}{E^\prime_\gamma} = \frac{3 L_\gamma}{4
E_\gamma \pi c^3 \gamma^3 \Delta t^2}
\end{eqnarray}
Following the arguments leading to Eqs.\,22, we obtain a rather large conversion of energy into pions
\begin{eqnarray}
f_{\pi} \simeq \frac{R^\prime}{\lambda_{p \gamma}} \simeq
\frac{L_{\gamma}}{E_{\gamma}}\frac{1}{\gamma^2 \Delta t} \frac{3
\sigma_\Delta \langle x_{p \rightarrow \pi} \rangle}{4\pi c^2} \simeq .35
\times \bigg(\frac{L_\gamma}{10^{45}\rm{erg/s}}\bigg) \bigg(\frac{10\,
\rm{eV}}{E_\gamma}\bigg)  \bigg(\frac{30}{\gamma}\bigg)^2
\bigg(\frac{1000\, \rm{sec}}{\Delta{t}}\bigg)
\end{eqnarray}
When $f_\pi$ approaches unity, pions will be absorbed before decaying into neutrinos, requiring the substitution of $f_\pi$ by $1-e^{-f_\pi}$.

For protons to photoproduce pions on photons with the ubiquitous UV photons of $\sim 10 \,$eV energy,
\begin{eqnarray}
E^\prime_p > \frac{m_\Delta^2-m_p^2}{4 E^\prime_\gamma}.
\end{eqnarray}
Therefore, in the observer's frame,
\begin{eqnarray}
E_p > 1.4 \times 10^{19} \rm{eV} \bigg(\frac{\gamma}{30}\bigg)^2
\bigg(\frac{10\, eV}{E_\gamma}\bigg).
\end{eqnarray}
If blazars are the sources of the highest energy cosmic rays, protons are accelerated to this energy and will generate accompanying neutrinos with energy
\begin{eqnarray}
E_\nu = \frac{1}{4} \langle x_{p \rightarrow \pi} \rangle  E_p  > 7
\times
10^{17} \rm{eV} \bigg(\frac{\gamma}{30}\bigg)^2 \bigg(\frac{10\,
\rm{eV}}{E_\gamma}\bigg).
\end{eqnarray}
The neutrino flux from blazars can be calculated in the same way as for GRB
\begin{eqnarray}
\phi_\nu \simeq \frac{c}{4 \pi}\frac{1}{E_\nu}\bigg( \frac{1}{2}
(1-e^{-f_\pi})\, t_H \frac{dE}{dt} \bigg) e^{(1-e^{-f_\pi})},
\end{eqnarray}
\begin{eqnarray}
\phi_\nu = 10^{-15} \rm{cm}^{-2} \rm{s}^{-1} \bigg( \frac{7 \times
10^{17}\rm{eV}}{E_\nu} \bigg) \bigg(\frac{t_H}{10 \rm{Gyr}} \bigg) \bigg(
\frac{dE/dt}{4 \times 10^{44}} \bigg) (1-e^{-f_\pi})e^{(1-e^{-f_\pi})},
\end{eqnarray}
using Eq.\,23 and Eq.\,24. This a flux on the order of $100\,\rm{km}^{-2} \rm{yr}^{-1}$ over $2\pi$ steradian. The number of
detected events is obtained from Eq.\,26,
\begin{eqnarray}
N_{\rm{events}} \sim \phi_\nu  P_{\nu \rightarrow \mu} \sim 10 \,
\rm{km}^{-2} \rm{yr}^{-1} \bigg( \frac{7 \times 10^{17}\rm{eV}}{E_\nu}
\bigg)^{1/2} \bigg(\frac{t_H}{10 \rm{Gyr}} \bigg) \bigg( \frac{dE/dt}{4
\times 10^{44} \rm{Mpc}^{-3} \rm{yr}^{-1}} \bigg)
\end{eqnarray}
for $f_\pi\simeq.35$.  For values of $\gamma$ varying from 10 to 100, the number of events varies from 1 to 70 events $\,\rm{km}^{-2}\rm{yr}^{-1}$, respectively. Observation in a kilometer-scale detector should be possible.

We did of course make the assumption that blazars accelerate protons to the very highest energies observed in the cosmic ray spectrum. One can consider an alternative mechanism for producing neutrinos in blazars that does not invoke protons of such high-energy.

In line-emitting blazers, external photons with energies in the keV-MeV energy range are known to exist. They are clustered in clouds of quasi-isotropic radiation. Protons in the jets of lower energy relative to those contemplated in the previous section, can photoproduce pions in interactions with these clouds\cite{dermer1,dermer2, external, external2}. Consider a target of external photons of energy near $E_{\gamma,\rm{ext}}$ with a luminosity $L_{\gamma,\rm{ext}}$. The fraction of proton energy transfered to pions is approximately given by
\begin{eqnarray}
f_{\pi, \rm{ext}} \simeq f_{\pi, \rm{int}}
\frac{L_{\gamma,\rm{ext}}}{L_{\gamma,\rm{int}}}
\frac{E_{\gamma,\rm{ext}}}{E_{\gamma,\rm{int}}}.
\end{eqnarray}
The neutrino energy threshold is
\begin{eqnarray}
E_\nu > 7 \times 10^{13} \rm{eV} \bigg(\frac{\gamma}{30}\bigg)^2
\bigg(\frac{100\, \rm{keV}}{E_{\gamma,\rm{ext}}}\bigg).
\end{eqnarray}
We will no longer relate the flux of protons to cosmic rays.  Instead we introduce the luminosity of protons, $L_p$ above pion production threshold.  This is a largely unknown parameter although it has been estimated to be on the order of 10\% of the total luminosity\cite{berez10}.  The proton energy needed to exceed the threshold of Eq.\,42 is
\begin{eqnarray}
E_p > 1.4 \times 10^{15} \bigg(\frac{\gamma}{30}\bigg)^2 \bigg(\frac{100
\, \rm{keV}}{E_{\gamma,\rm{ext}}}\bigg).
\end{eqnarray}
The neutrino flux can be calculated as a function of $L_p$
\begin{eqnarray}
\Phi_\nu \simeq \frac{  \frac{1}{2} \langle x_{p \rightarrow \pi} \rangle
L_p f_{\pi, \rm{ext}} \Delta t   }{ E_\nu 4 \pi D^2},
\end{eqnarray}
where $\Delta t$ is the duration of a blazar flare.  This reduces to
\begin{eqnarray}
\Phi_\nu \sim 4 \times 10^4 \,{\rm{km}}^{-2} \bigg(\frac{f_{\pi,
\rm{ext}}}{.5}\bigg) \bigg( \frac{L_p}{10^{45}\,\rm{erg/s}}\bigg) \bigg(
\frac{\Delta
t}{1000\,\rm{sec}}\bigg) \bigg(\frac{1000\,\rm{Mpc}}{D}\bigg)^2
\bigg(\frac{30}{\gamma} \bigg)^2
\bigg(\frac{E_{\gamma,\rm{ext}}}{100\,\rm{keV}}\bigg)
\end{eqnarray}
for a fifteen minute flare. Using Eq.\,26, the event rate of TeV neutrinos is
\begin{eqnarray}
N_{\rm{events}} \sim \phi_\nu  P_{\nu \rightarrow \mu} \sim 2 \,
{\rm{km}}^{-2}  \bigg(\frac{f_{\pi, \rm{ext}}}{.5}\bigg) \bigg(
\frac{L_p}{10^{45}\,\rm{erg/s}}\bigg) \bigg( \frac{\Delta
t}{1000\,\rm{sec}}\bigg) \bigg(\frac{1000\,\rm{Mpc}}{D}\bigg)^2
\bigg(\frac{30}{\gamma} \bigg)^{2/5}
\bigg(\frac{E_{\gamma,\rm{ext}}}{100\,\rm{keV}}\bigg)^{1/5}.
\end{eqnarray}
Note that this result is for a typical, but fairly distant source.  A nearby line-emitting blazar could be a
strong candidate for neutrino observation. Considering the more than 60 blazars which have been observed, the total
flux may generate conservatively  tens or, optimistically, hundreds of TeV-PeV neutrino events per year in a kilometer scale neutrino telescope such as IceCube.  The upper range of this estimate can be explored by the AMANDA experiment at the level of a few events per year.

Blazar neutrino searches should be able to find incontrovertible evidence for cosmic ray acceleration in active galaxies, or, alternatively, challenge the possibility that AGN are the sources of the highest energy cosmic rays.

The case for doing neutrino astronomy is compelling, the challenge has been to deliver the technology to build the instrumentation for a neutrino detector that is required, unfortunately, to have kilometer-scale dimensions. We discuss this next.

\section{High Energy Neutrino Telescopes: Methodologies}

The construction of neutrino telescopes is motivated by their discovery potential in astronomy, astrophysics, cosmology and particle physics.  To maximize this potential, one must design an instrument with the largest possible effective telescope area to overcome the small neutrino cross section with matter, and the best possible angular and energy resolution to address the wide diversity of possible signals. While the smaller first-generation detectors have been optimized to detect secondary muons initiated by $\nu_{\mu}$, kilometer-scale neutrino observatories will detect all neutrino flavors over a wide range of energies.

We will first review the methods to identify neutrinos, measure their energy and identify their flavor. Next we will briefly review a variety of projects launched with the goal to identify cosmic neutrinos sources beyond the Sun. 

\subsection{Detection Technique}

High energy neutrinos are detected by observing the Cherenkov radiation from secondary particles produced by neutrinos interacting inside large volumes of highly transparent ice or water instrumented with a lattice of photomultiplier tubes (PMT). For simplicity, assume an instrumented cubic volume of side $L$; see Fig.\,5. (Also assume, for simplicity, that the neutrino direction is perpendicular to a side of the cube; for a realistic detector the exact geometry has to be taken into account as well as the arrival directions of the neutrinos; more later.) 

\begin{figure}[!h]
\centering\leavevmode
\includegraphics[width=4.25in]{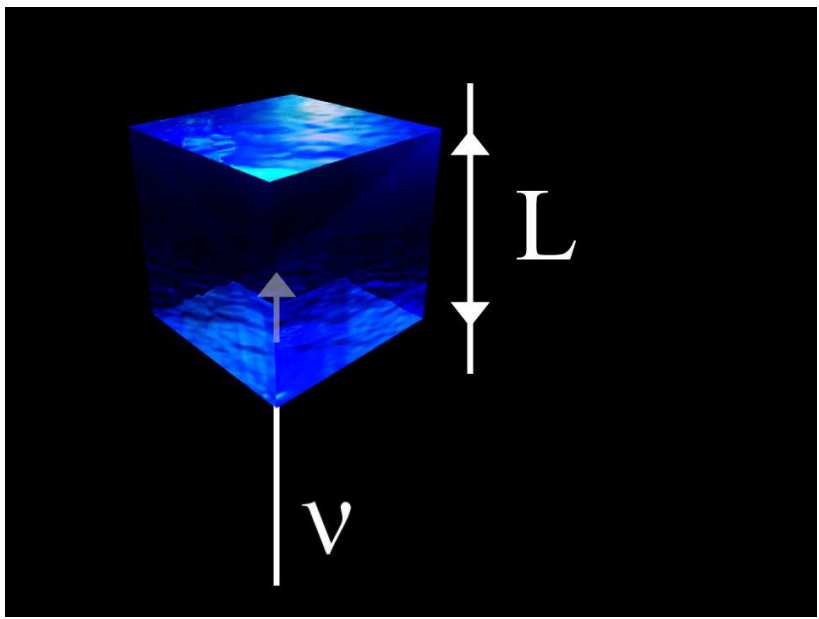}
\caption{A neutrino interacts in a cube of instrumented ice of side L.}
\end{figure}

To a first approximation, a neutrino incident on a side of area $L^2$ will be detected provided it interacts within the detector volume, i.e. within the instrumented distance $L$. That probability is
\begin{equation}
P = 1 - \exp(-L/\lambda_{\nu}) \simeq L/ \lambda_{\nu} \,,
\end{equation}
with $\lambda_{\nu} = (\rho N_A \sigma_{\nu})^{-1}$. Here $\rho$ is the density of the ice or water, $N_A$ Avogadro's number and $\sigma_{\nu}$ the neutrino cross section.
A neutrino flux $\phi$ (neutrinos per $cm^2$ per second) crossing a detector with cross sectional area $A\ (\simeq L^2)$ facing the incident beam, will produce
\begin{equation}
N = AT P \phi
\end{equation}
events after a time T. In practice, the quantities $A$, $P$ and $\phi$ depend on the neutrino energy and $N$ is obtained by a convolution over neutrino energy above the detector threshold.

The "effective" telescope area A is not strictly equal to the geometric cross section of the instrumented volume facing the incoming neutrino because even neutrinos interacting outside the instrumented volume, may produce a sufficient amount of light inside the detector to be detected. In practice, A is therefore determined as a function of the incident neutrino direction by simulation of the full detector, including the trigger.

The formalism presented applies to electron neutrinos. For muon neutrinos, any neutrino producing a secondary muon that reaches the detector (and has sufficient energy to trigger it), will be detected. Because the muon travels kilometers at TeV energies and tens of kilometers at EeV energy, neutrinos can be detected far outside the instrumented volume; the probability is obtained by the substitution
\begin{equation}
L \rightarrow \lambda_{\mu} \,,
\end{equation}
therefore,
\begin{equation}
P = \lambda_{\mu}/\lambda_{\nu} \,.
\end{equation}
Here $ \lambda_{\mu}$ is the range of the muon determined by its energy losses. The complete expression for the flux of $\nu_\mu$-induced muons at the detector is given by a convolution of the neutrino spectrum $\phi$ with the probability P to produce a muon reaching the detector\cite{PR}:
\begin{equation}
\phi_\mu(E_\mu^{\rm min},\theta) =
\int_{E_\mu^{\rm min}}\,P(E_\nu,E_\mu^{\rm min})\,
\exp[-\sigma_{\rm tot}(E_\nu)\,N_A\,X(\theta)]\,\phi(E_\nu,\theta).
\label{N_mu}
\end{equation}
Here $E_\mu^{\rm min}$ is the detector threshold for muons. The exponential factor here accounts for absorption of neutrinos along the chord of the Earth, $X(\theta)$.  Absorption becomes
important for $\sigma_\nu(E_\nu)\gtrsim 10^{-33}$\,cm$^2$ or $E_\nu\gtrsim 10^7$~GeV. On the other hand, for back-of-the-envelope calculations, the $P$-function can be approximated by
\begin{eqnarray}
P \simeq 1.3 \times 10^{-6} E^{2.2}  && \mbox{for $E = 10^{-3}$--1 TeV} \,, \\
    \simeq 1.3 \times 10^{-6} E^{0.8}  && \mbox{for $E= 1$--$10^3$ TeV} \,.
\end{eqnarray}
At EeV energy the increase is reduced to only $E^{0.4}$.
  
Similar arguments apply to the detection of a tau neutrino that will be detected provided the tau lepton it produces reaches the instrumented volume within its lifetime. Therefore
\begin{equation}
L \rightarrow \gamma c \tau = E/m c \tau\,,
\end{equation}
where $m$, $\tau$ and $E$ are the mass, lifetime and energy of the tau, respectively. $\gamma c \tau$ for tau leptons grows linearly with energy and exceeds the range of the muon around 1\,EeV.

The large cross sections of neutrinos, the long range of the muon and the long lifetime of the tau make the construction of high energy neutrino detectors of kilometer-scale dimension possible. Muons and tau neutrinos can be detected over volumes of ice and water larger than those instrumented with PMT. For illustration we show in Fig.\,6a the effective volume for electromagnetic showers of IceCube. Note that at high energy the  ``effective" volume exceeds the 1\,km$^3$ volume instrumented. Also the effective area for the detection of $\nu_{\mu}$ exceeds 1\,km$^2$, the cross section of instrumented ice facing a neutrino beam; see Fig.\,6b.

\begin{figure}[!h]
\centering
\subfigure[]
{
	\label{fig:sub:a}
	\includegraphics[width=8cm]{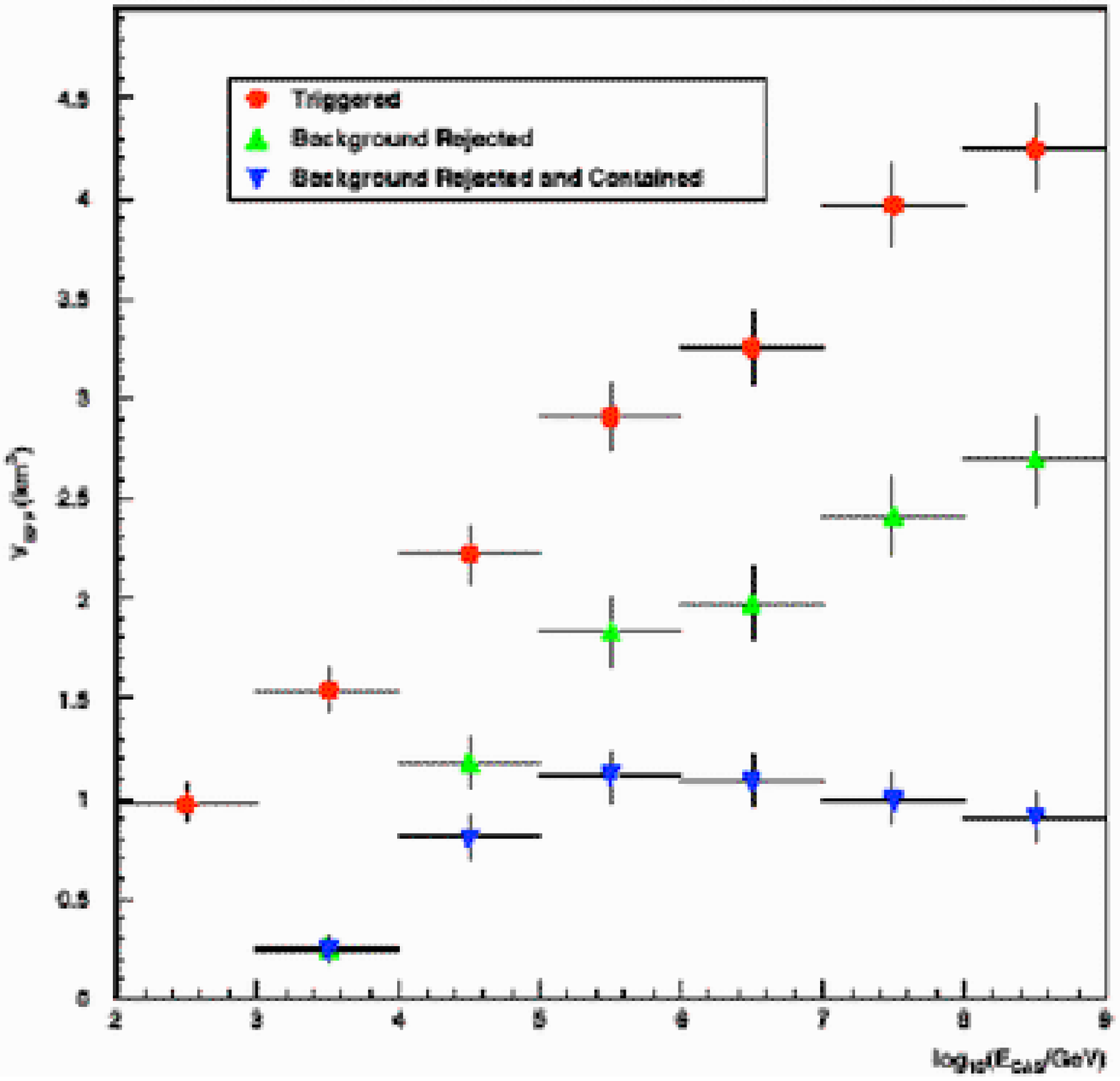}
}
\hspace{1cm}
\subfigure[]
{
	\label{fig:sub:b}
	\includegraphics[width=8cm]{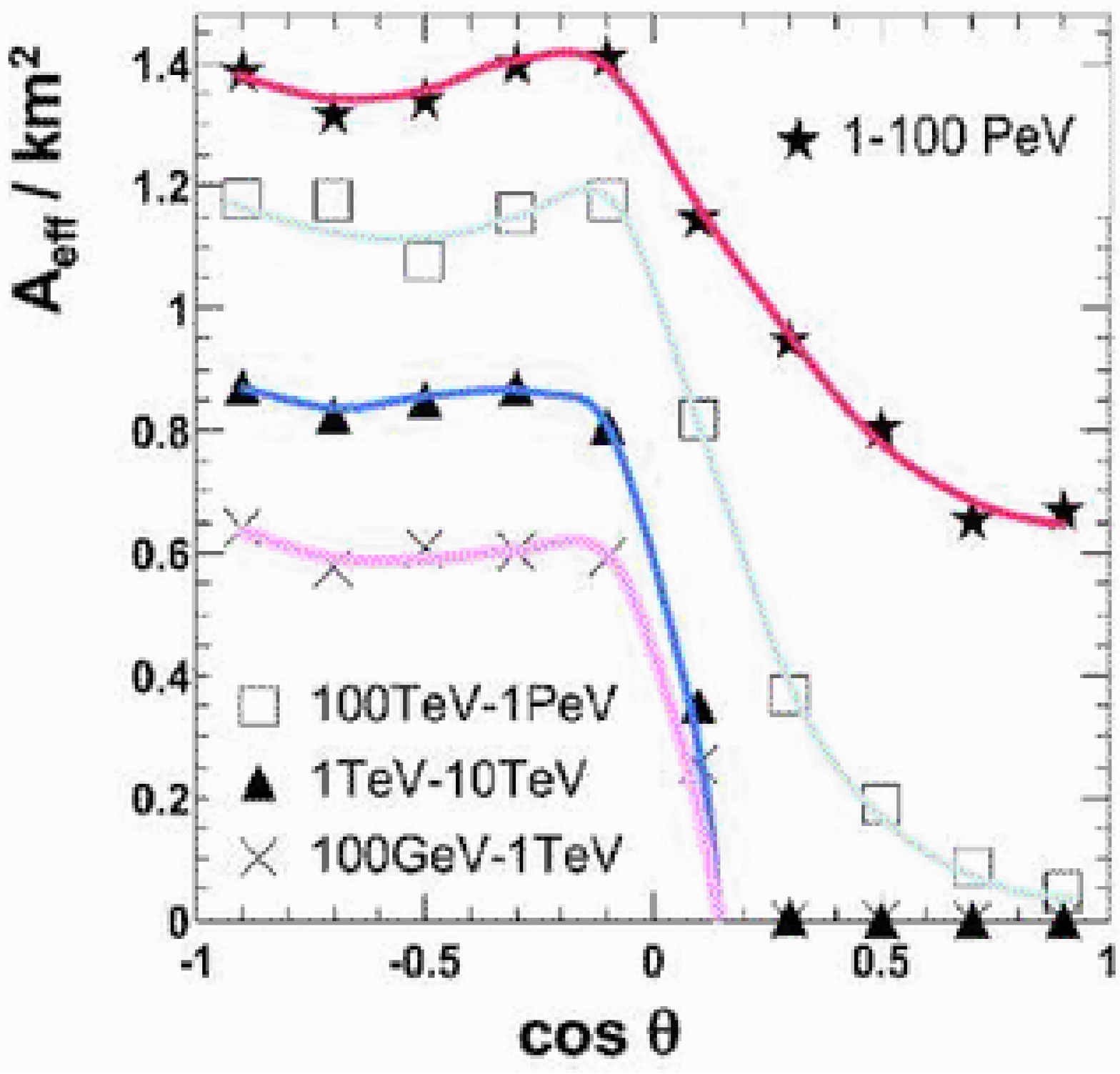}
}
\caption{ Fig.6(a) illustrates the effective volume of IceCube for the detection of showers initiated by neutrinos of electron or tau flavor. Fig.6(b) illustrates the effective area of IceCube for the detection of neutrinos of muon flavor. The performance of IceCube has been simulated with AMANDA analogue signals rather than with the superior digital signals of IceCube. IceCube performance should be superior to what is shown.}
\label{Caption for the whole figure}
\end{figure}

For "realistic" calculations, even Eq. 51 is simplified. The complete expression to compute, for instance, the expected number of $\nu_\mu$ induced events is given by
\begin{eqnarray}
N^{\nu_\mu}_{\rm ev}
&=& T \int^{1}_{-1} d\cos\theta\,  
\int^\infty_{l'min} dl\,
\int_{m_\mu}^\infty dE_\mu^{\rm fin}\,
\int_{E_\mu^{\rm fin}}^\infty dE_\mu^0\, 
\int_{E_\mu^0}^\infty dE_\nu \\ \nonumber
&&\frac{d\phi_{\nu_\mu}}{dE_\nu d\cos\theta}(E_\nu,\cos\theta)
\frac{d\sigma^\mu_{CC}}{dE_\mu^0}(E_\nu,E_\mu^0)\, n_T\, 
F(E^0_\mu,E_\mu^{\rm fin},l)\, A^0_{eff}\, \, .
\label{eq:numuevents}
\end{eqnarray}
$\frac{d\phi_{\nu_\mu}}{dE_\nu d\cos\theta}$ is the differential muon
neutrino neutrino flux in the vicinity of the detector after propagation through the Earth matter. Because of the high energies of the neutrinos,
their oscillations, propagation in the Earth and regeneration
due to  $\tau$ decay must be treated in a coherent way\cite{GHM}. $\frac{d\sigma^\mu_{CC}}{dE_\mu^0}(E_\nu,E_\mu^0)$ is the differential charged current interaction cross section producing a muon of energy $E_\mu^0$ and
$n_T$ is the number density of nucleons in the matter surrounding the
detector and $T$ is the exposure time of the detector. After production with energy $E_\mu^0$, the muon ranges out in the
rock and in the ice surrounding the detector and looses energy.  $F(E^0_\mu,E_\mu^{\rm fin},l)$ the function that describes
the energy spectrum of the muons arriving at the detector.  Thus
$F(E^0_\mu,E_\mu^{\rm fin},l)$ represents the probability that a muon
produced with energy $E_\mu^0$ arrives at the detector with energy
$E_\mu^{\rm fin}$ after traveling a distance $l$. The function $F(E^0_\mu,E_\mu^{\rm fin},l)$ is computed by propagating the muons to
the detector taking into account energy losses due to ionization,
bremsstrahlung, $e^+e^-$ pair production and nuclear interactions; see for instance Ref.~\cite{ls}.

Equivalently, muon events arise from $\bar\nu_\mu$ interactions. They are calculated from an equation similar to Eq.(\ref{eq:numuevents}) as are the event rates induced by neutrino fluxes of electron and tau flavor.

Detection of all neutrino flavor has become important for two reasons: neutrino oscillations and tau neutrino ``regeneration" through the Earth. The generic cosmic accelerator is believed to produce neutrinos from the decay of pions in the flux ratio $\nu_e:\nu_\mu:\nu_\tau = 1:2:0$.  This is also the admixture expected for the atmospheric neutrino beam below 10 GeV where the muons decay. With neutrino oscillations, however, at the detection point this ratio becomes $1:1:1$ because approximately one half of the muon neutrinos convert to tau flavor over cosmic baselines.  This represents and advantage for neutrino telescopes because the $\nu_\tau$, unlike the $\nu_e$ and $\nu_\mu$, are not absorbed in the Earth due to the charged current regeneration effect:  $\nu_\tau$'s with energies exceeding 1~PeV pass through the Earth and emerge with an energy of roughly 1~PeV. The mechanism is simple\cite{H&S}. A $\nu_{\tau}$ interacting in the Earth will produce another $\nu_{\tau}$ of lower energy, either directly in a neutral current interaction or via the decay of a tau lepton produced in a charged current interaction. High energy $\nu_{\tau}$ will thus cascade down to PeV energy where the Earth is transparent; in other words, the Earth cannot lower their energy below the hundreds of GeV thresholds of the detector.

\subsection{Neutrino Flavors}

Kilometer-scale neutrino telescopes are designed to detect neutrinos of all energies above ${\sim}100$\,GeV. They are designed to measure the energy of the neutrinos in the range 100\,GeV--100\,EeV and to identify the neutrino flavor over as large an energy range as possible.

\subsubsection{Electron Neutrinos} 

At the most basic level, neutrino telescopes detect the secondary particle showers initiated by neutrinos of all flavors as well as the leading secondary muon tracks initiated by $\nu_\mu$ only-- see next section. High energy electron neutrinos deposit 0.5-0.8\% of their energy into an electromagnetic shower initiated by the leading final state electron. The rest of the energy goes into the fragments of the target that produce a second subdominant shower. Because the size of the shower, of order meters in ice or water, is small compared to the spacing of the PMTs, it represents, to a good approximation, a point source of Cherenkov photons radiated by the shower particles. These trigger the PMT at the single photoelectron level over a spherical volume whose radius scales linearly with the shower energy. For ice, the radius is 130\,m at 10\,TeV and 460\,m at 10\,EeV, i.e. the shower radius grows by just over 50\,m per decade in energy; see Fig.\,7. 

\begin{figure}[!h]
\centering\leavevmode
\includegraphics[width=5in]{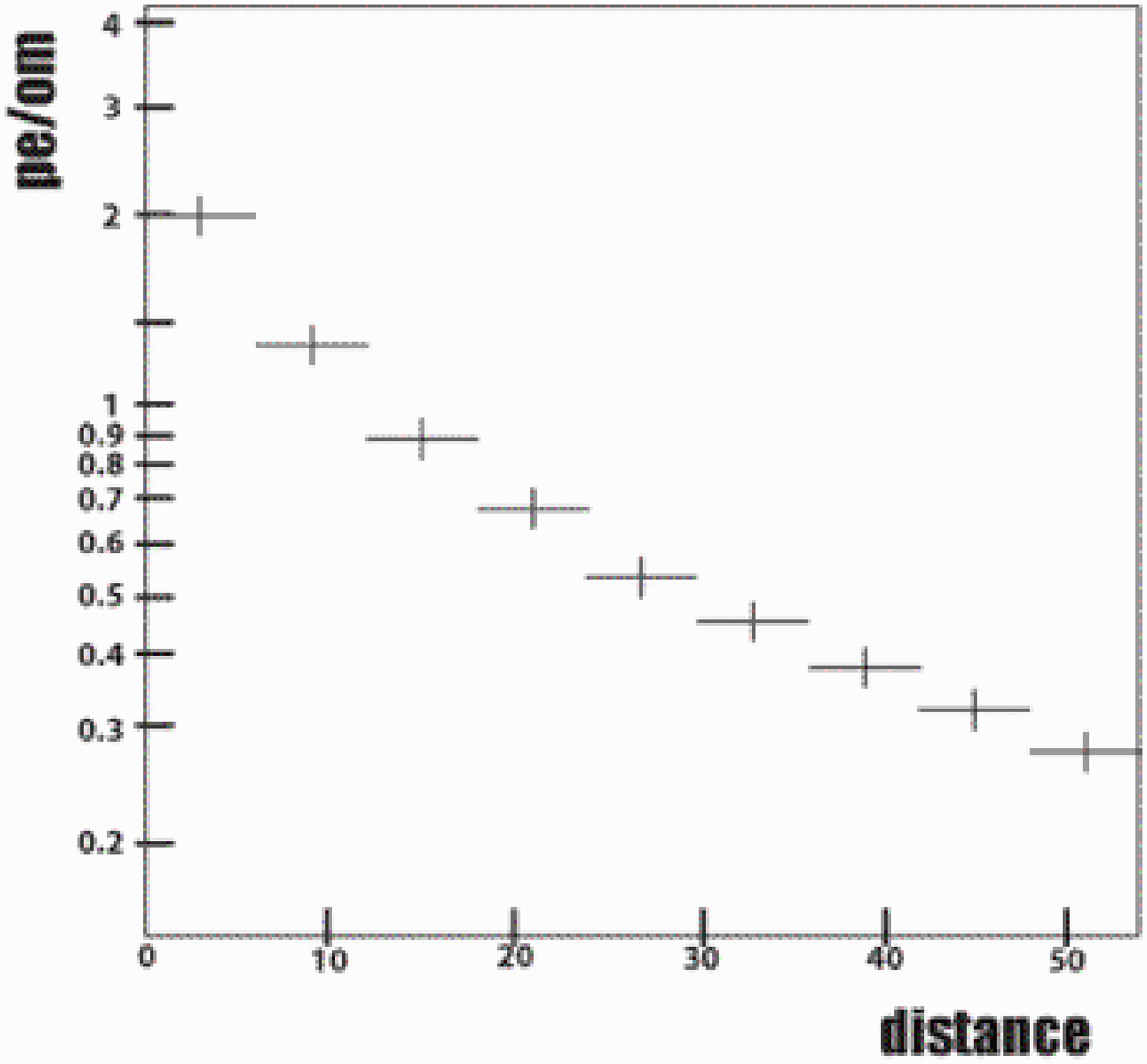}
\caption{Distance over which a 10 inch photomultiplier, set at a photoelectron threshold (pe), detects a minimum ionizing muon in ice. }
\end{figure}

The measurement of the radius of the sphere in the lattice of PMT determines the energy and turns neutrino telescopes into total energy calorimeters. Note that a contained ``direct hit" by a 10 EeV neutrino will not saturate a km$^3$ detector volume. So, even for GZK neutrinos, neutrino telescopes do not saturate. 

Because the shower and its accompanying Cherenkov lightpool are not totally symmetric but elongated in the direction of the leading electron (and incident neutrino), its direction can be reconstructed. Pointing is however far inferior to what can be achieved for muon neutrinos and estimated to be precise to ${\sim}10$\ degrees only. On the other hand, compared to $\nu_{\mu}$, energy reconstruction is superior and the background of atmospheric neutrinos is significantly reduced. At higher energies muons, the source of atmospheric $\nu_{e}$ , no longer decay.

\subsubsection{Muon Neutrinos}

Secondary muons initiated by muon neutrinos range out over kilometers at TeV energy, to tens of kilometers at EeV energy, generating showers along their track by bremsstrahlung, pair production at TeV energy and photonuclear interactions. These are the sources of Cherenkov radiation and are detected in exactly the same way as the leading electron neutrino in the previous section. As the energy of the muon degrades along its track, the energy of the secondary showers diminishes and the distance from the track over which the associated Cherenkov light can trigger a PMT becomes smaller. The geometry of the lightpool surrounding the muon track over which single photo-electron are produced is therefore a kilometer-long cone with gradually decreasing radius.

High energy muons lose energy catastrophically according to
\begin{equation}
\frac{dE}{dX}=-\alpha - \beta E \, ,
\end{equation}
where $\alpha=2.0 \times 10^{-6}~{\rm TeV}\, {\rm cm}^2/{\rm g}$ and $\beta=4.2 \times
10^{-6}~{\rm cm}^2/{\rm g}$. The distance a muon travels before its energy drops below some energy threshold, $E^{\rm thr}_\mu$, called the muon range is then given by
\begin{equation}  
\lambda_\mu = \frac{1}{\beta} \ln \left[ 
\frac{\alpha + \beta E_\mu}{\alpha + \beta E^{\rm thr}_\mu} \right] \,.
\label{murange}
\end{equation}

In the first kilometer a high energy muon typically loses energy in a couple of showers of one tenth its energy. So the initial size of the cone is the radius of a shower with 10\% of the muon energy, e.g.\ 130\,m for a 100\,TeV muon. Near the end of its range the muon becomes minimum ionizing emitting light that creates single photoelectron signals at a distance of just over 10\,m from the track. For 0.3 photoelectrons, the standard PMT threshold setting, this distance reaches 45\,m.

Note however that the energy measurement is indirect unlike the case for showers. Because of the stochastic nature of muon energyloss, the logarithm of the energy is measured. Also, although at PeV energy and above, muons have ranges of tens of kilometers, greatly enhancing their detectability, the initial energy of the event cannot always be measured. A muon can be produced at one energy, travel several kilometers, and be detected with much less energy; see Eq. 55.

\subsubsection{Tau Neutrinos} 

Interest in the detection of $\tau$ neutrinos is motivated by the discovery that half of the muon neutrinos convert to tau neutrinos over cosmic distances. Production of $\nu_\tau$ in hadronic interactions or photoproduction in the heavenly bean dump is suppressed relative to $\nu_e$ and $\nu_\mu$ by
some five orders of magnitude.  In the absence of oscillations, $\nu_\tau$ of astrophysical origin would have been virtually undetectable. Now they form 1/3 of the beam and have the additional advantage not to be absorbed by the Earth; although they may loose energy, they always reach the detector.

In a kilometer-scale neutrino detector the flavor of tau neutrinos of sufficiently high energy can be identified in several ways.  Perhaps the most striking signature is the characteristic double bang event\cite{PR} in which the production and decay of a $\tau$ lepton are detected as two separated showers inside the detector. It may
also be possible to identify ``lollipop'' events in which a $\nu_\tau$ creates a long minimum-ionizing track that enters
the detector and ends in a huge burst when the $\tau$ lepton decays to a final state with hadrons or an electron.  The parent $\tau$ track can be identified by the reduced catastrophic energy loss compared to a muon of similar energy. In other words, the large energy of the shower cannot be generated by a muon that should have revealed its muon flavor by abundant radiation along the early part of the track.

The efficiency for a kilometer-scale detector to identify double-bang events can be estimated as follows. In a charged current $\nu_\tau$ deep inelastic scattering interaction with a nucleus, a $\tau$ lepton of energy $(1-y)E_{\nu_\tau}$ is produced as well as a hadronic shower of energy $y E_{\nu_\tau}$ initiated by the fragmentation of the target. Here $y$ is the fraction of energy transferred to the hadronic vertex in the interaction. Before decaying, the $\tau$ lepton travels on average a distance $R_\tau$ given by:
\begin{equation}
\lambda_\tau={E_\tau\over m_\tau} ct_0={(1-y)E_{\nu_\tau}\over m_\tau}c\tau
\label{taurange}
\end{equation}
where $E_\tau$ and $m_\tau$ are the energy and mass of the $\tau$ respectively and $\tau$ is its lifetime at rest.  The decay produces another $\nu_\tau$ and an electromagnetic or hadronic shower $\sim82\%$ of the times.  Assuming a detector of dimension $L$, there are several conditions that have to be fulfilled for  identification of a double bang event:

\begin{itemize}

\item The $\nu_\tau$ has to interact through a charged current interaction producing a hadronic shower contained inside
the instrumented volume ($\sim L^3$).

\item The $\tau$ lepton must decay inside the detector to a final state
that produces an electromagnetic or hadronic shower which has to
be contained inside the device.

\item $\lambda_\tau$ has to be sufficiently large for the two showers to be clearly separated.

\item The showers must be sufficiently energetic to trigger the detector.

\end{itemize}

In the vicinity of 10\,PeV the probability to detect and identify a $\nu_\tau$ as a double-bang is only 10\% of that for detecting a $\nu_\mu$ of the same energy. At lower and higher energies the likelihood of detecting a double-bang falls rapidly.

\section{High Energy Neutrino Telescopes: Status}

In this section we discuss the first first-generation detector AMANDA with effective telescope area of $1\sim8\times10^4m^2$, depending on the science. We will subsequently briefly review the efforts to build a similar detector in the Mediterranean. For neutrino astronomy to become a viable science, several projects will have to succeed. Astronomy, whether in the optical or in any other wave-band, thrives on a diversity of complementary instruments, not on ``a single best instrument".

\subsection{Neutrino Telescopes: First ``Light"}

While it has been realized for many decades that the case for neutrino astronomy is compelling, the challenge has been to develop a reliable, expandable and affordable detector technology to build the kilometer-scale telescopes required to do the science. Conceptually, the technique is simple. In the case of a high-energy muon neutrino, for instance, the neutrino interacts with a hydrogen or oxygen nucleus in deep ocean water and produces a muon travelling in nearly the same direction as the neutrino. The Cerenkov light emitted along the muon's kilometer-long trajectory is detected by a lattice of photomultiplier tubes deployed on strings at depth shielded from radiation. The orientation of the Cerenkov cone reveals the neutrino direction.

The AMANDA detector, using natural 1 mile-deep Antarctic ice as a Cerenkov detector, has operated for more than 5 years in its final configuration of 680 optical modules on 19 strings. The detector is in steady operation collecting roughly $7 \sim 10$ neutrinos per day using fast on-line analysis software. The lower number will yield a background-free sample all the way to the horizon. AMANDA's performance has been calibrated by reconstructing muons produced by atmospheric muon neutrinos\cite{nature}.

\begin{figure}[!h]
\centering\leavevmode
\includegraphics[width=5in]{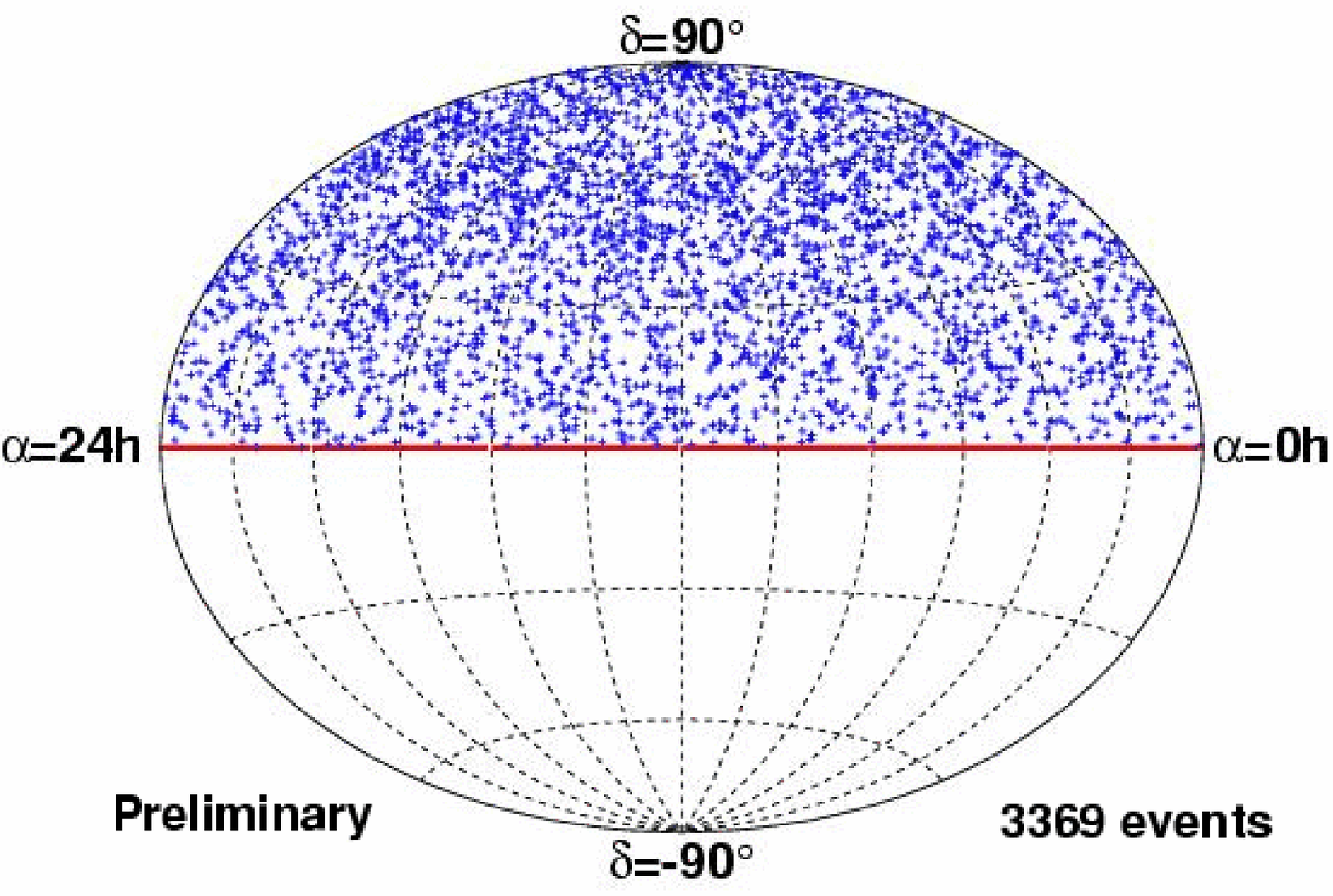}
\caption[]{Skymap showing declination and right ascension of neutrinos detected by the AMANDA\,II detector during four Antarctic winters of operation in 2000-2003.\label{fig:skyplot}}
\end{figure}

Using the first 4 years of AMANDA\,II data, the AMANDA collaboration is performing a search for  the emission of muon neutrinos from spatially localized directions in the northern sky. Only the year 2000 data has been published\cite{HS}. The skyplot for all 4 years is shown in Fig.~\ref{fig:skyplot}. The 90\% upper limits on the neutrino fluency of point sources are at the level of $6 \times 10^{-8}\rm\, GeV \,cm^{-2}\, s^{-1}$ or $10^{-10}\rm \,erg\ cm^{-2}\,s^{-1}$, averaged over declination . This corresponds to a flux of $6 \times 10^{-9}\rm\, cm^{-2}\, s^{-1}$ integrated above 10\,GeV assuming a $E^{-2}$ energy spectrum typical for shock acceleration of particles in high energy sources. The most significant excess is 3.4\,$\sigma$ from the Crab with a probability of close to 10\% given the trial factor for 33 sources searched. IceCube is needed to make conclusive observations of sources.

The AMANDA\,II detector has reached a high-energy effective telescope area of 25,000$\sim$40,000\,m$^2$, depending on declination. This represents an interesting milestone\cite{alvarezhalzen}: known TeV gamma ray sources, such as the active galaxies Markarian 501 and 421, should be observed in neutrinos if the number of gamma rays and neutrinos emitted are roughly equal as expected from cosmic ray accelerators producing pions\cite{alvarezhalzen}. Therefore AMANDA must detect the observed TeV photon sources soon, or, its observation will exclude them as the sources of cosmic rays.

\subsection{Mediterranean Telescopes}

Below PeV energy, South Pole neutrino telescopes do not cover the Southern sky, which is obscured by the large flux of cosmic ray muons. This and the obvious need for more than one telescope --- accelerator experiments have clearly demonstrated the value of multiple detectors --- provide compelling arguments for deploying northern detectors. With the first observation of neutrinos by a detector in Lake Baikal with a telescope area of 2500\,m$^2$ for TeV muons\cite{baikal} and after extensive R\&D efforts by both the ANTARES\cite{antares} and NESTOR\cite{nestor} collaborations in the Mediterranean, there is optimism that the technological challenges to build neutrino telescopes in deep sea water have been met.  Both Mediterranean collaborations have demonstrated their capability to deploy and retrieve optical sensors, and have reconstructed down-going muons with optical modules deployed for R\&D tests.

The ANTARES neutrino telescope is under construction at a 2400\,m deep Mediterranean site off Toulon, France. It will consist of 12 strings, each equipped with 75 optical sensors mounted in 25 triplets. The detector performance has been fully simulated\cite{antares} with the following results: a sensitivity after one year to point sources of $0.4-5 \times 10^{-15}\rm\, cm^{-2}\, s^{-1}$ (note that this is the flux of secondary muons, not neutrinos) and to a diffuse flux of $0.9 \times 10^{-7}\rm\, GeV \,cm^{-2}\, s^{-1}$ above 50\,TeV. As usual, an $E^{-2}$ spectrum has been assumed for the signal. AMANDA\,II data have reached similar point source limits ($0.6 \times 10^{-15}\rm\, cm^{-2}\, s^{-1}\,sr^{-1}$) using 4 Antarctic winters, or about 1000 days, of data\cite{HS}). This value depends weakly on declination. Also the diffuse limits reached in the absence of a signal are comparable\cite{hill}. We have summarized the sensitivity of both experiments in Table\,2 where they are also compared to the sensitivity of IceCube.

\begin{table}[h]
\caption{}
\tabcolsep1.25em
\def\arraystretch{1.5}
\begin{center}
\begin{tabular}{|c|c|c|c|}
\hline
& \bf  IceCube& \bf AMANDA-II$^*$& \bf ANTARES\\
\hline
\bf \# of PMTs& 4800 / 10 inch& 600 / 8 inch& 900 / 10 inch\\
\hline
\parbox{6em}{{ \bf Point source sensitivity} (muons/year)}& $6\times 10^{-17}\,\rm cm^{-2}\,\rm s^{-1}$&
\parbox{8.5em}{\centering $1.6\times 10^{-15}\,\rm cm^{-2}\, s^{-1}$ weakly dependent\\ on oscillations}&
\parbox{9.5em}{\centering 0.4--$5\times 10^{-15}\rm\, cm^{-2}\,s^{-1}$ depending\\ on declination}\\[.2in]
\hline
\parbox{6em}{{\bf diffuse limit$^\dagger$}\\ (neutrinos/year)}& 
\parbox{7.5em}{\centering 3--$12\times 10^{-9}\,\rm GeV\break cm^{-2}\,s^{-1}\,sr^{-1}$}&
\parbox{5.5em}{\centering $2\times 10^{-7}\rm\,GeV\break cm^{-2}\, s^{-1}\,sr^{-1}$}&
\parbox{6.5em}{\centering $0.8\times 10^{-7}\rm\,GeV\break cm^{-2}\,s^{-1}\,sr^{-1}$}\\[.1in]
\hline
\multicolumn{4}{l}{$^*$includes systematic errors}\\
\multicolumn{4}{l}{$^\dagger $depends on assumption for background from atmospheric neutrinos from charm}
\end{tabular}
\end{center}
\end{table}

Given that AMANDA and ANTARES operate at similar depths and have similar total photocathode area (AMANDA\,II is actually a factor of 2 smaller with 667 8-inch versus 900 10-inch photomultipliers for Antares), the above comparison provides us with a first glimpse at the complex question of the relative merits of water and ice as a Cherenkov medium. The conclusion seems to be that, despite differences in optics and in the background counting rates of the photomultipliers, the telescope sensitivity is approximately the same for equal photocathode area. The comparison is summarized in Table~1 where we have tabulated the sensitivity of AMANDA and Antares to points sources and to a diffuse flux of neutrinos. At this time, in the absence of a discovery, it is the sensitivity and not the area or angular resolution that represents the relevant quantity of merit. In the same context, the NEMO collaboration has done the interesting exercise of simulating the IceCube detector (augmented from 4800 to 5600 optical modules; see next section) in water rather than ice. One finds a slightly reduced sensitivity in water, probably not significant within errors and at no energy larger than 50\%\cite{emigneco}. Note that in several years of operation a kilometer-scale detector like IceCube can improve the sensitivity of first-generation telescopes by two orders of magnitude.

\subsection{Kilometer-scale Neutrino Observatories}

The baseline design of kilometer-scale neutrino detectors maximizes sensitivity to $\nu_\mu$-induced muons with energy above hundreds of GeV, where the acceptance is enhanced by the increasing neutrino cross section and muon range and the Earth is still largely transparent to neutrinos. The mean-free path of a $\nu_\mu$ becomes smaller than the diameter of the Earth above 70\,TeV --- above this energy neutrinos can only reach the detector from angles closer to the horizon. Good identification of other neutrino flavors becomes a priority, especially because $\nu_\tau$ are not absorbed by the Earth. Good angular resolution is required to distinguish possible point sources from background, while energy resolution is needed to enhance the signal from astrophysical sources, which are expected to have flatter energy spectra than the background atmospheric neutrinos.

Overall, AMANDA represents a proof of concept for the kilometer-scale neutrino observatory, IceCube\cite{ice3}, now under construction. IceCube will consist of 80 kilometer-length strings, each instrumented with 60 10-inch photomultipliers spaced by 17~m. The deepest module is 2.4~km below the surface. The strings are
arranged at the apexes of equilateral triangles 125\,m on a side. The instrumented (not effective!) detector volume is a cubic kilometer. A surface air shower detector, IceTop, consisting of 160 Auger-style Cherenkov detectors deployed over 1\,km$^{2}$ above IceCube, augments the deep-ice component by providing a tool for calibration, background rejection and air-shower physics, as illustrated in Fig. 9.

\begin{figure}[!h]
\centering\leavevmode
\includegraphics[width=5in]{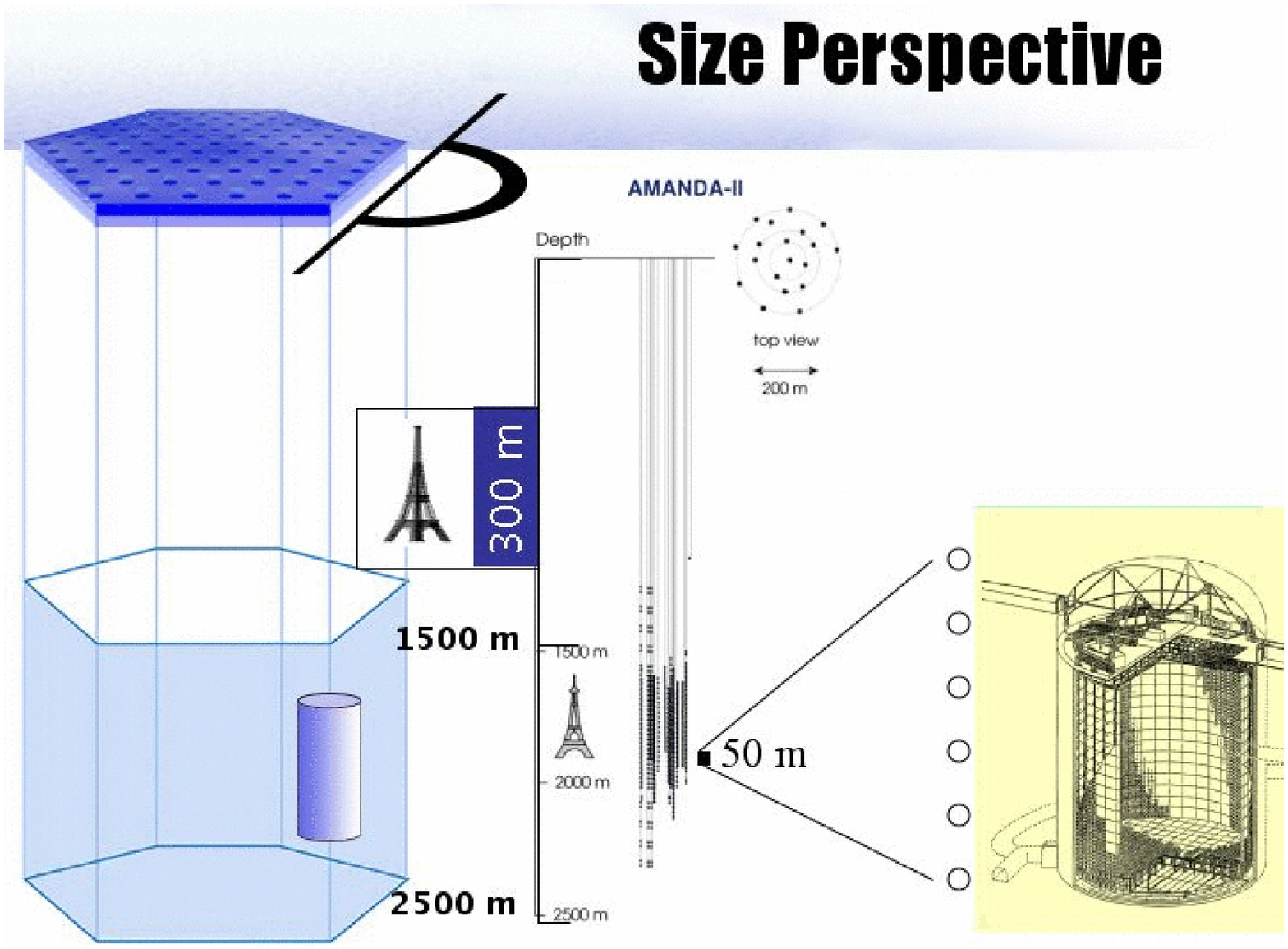}
\caption{Relative sizes of the IceCube, AMANDA, and Superkamiokande neutrino detectors. AMANDA will be operated as a lower threshold subsystem of IceCube. As the size of the detector grows, so does the threshold energy of neutrinos detected.}
\end{figure}

The transmission of analogue photomultiplier signals from the deep ice to the surface, used in AMANDA, has been abandoned. The photomultiplier signals will be captured and digitized inside the optical module.  The digitized signals are given a global time stamp with a precision of $<10$\,ns and transmitted to the surface.  The digital messages are sent to a string processor, a global event trigger and an event builder.  

Construction of the detector commenced in the Austral summer
 of 2004/2005 with the assembly of the hot water drill and the deployment of the first 60 digital optical modules. It will continue for 6 years, possibly less.  The growing detector will take data during construction, with each string coming online within days of deployment. The data streams of IceCube, and AMANDA\,II, embedded inside IceCube,  will be merged off-line using GPS timestamps.

IceCube will offer advantages over AMANDA\,II beyond its larger size: it will have a higher efficiency and superior angular resolution in reconstructing tracks, map showers from electron- and tau-neutrinos (events where both the production and decay of a $\tau$ produced by a $\nu_{\tau}$ can be identified) and, most
importantly, measure neutrino energy. Simulations, benchmarked by AMANDA data, indicate that the direction of muons can be determined with sub-degree accuracy and
their energy measured to better than 30\% in the logarithm of the energy. The direction of showers will be reconstructed to better than 10$^\circ$ above 10\,TeV and the response in energy is linear and better than 20\%. Energy resolution is critical because, once one establishes that the energy exceeds 1\,PeV, there is no atmospheric muon or neutrino background in a kilometer-square detector and full sky coverage of the telescope is achieved. The background counting rate of IceCube signals is expected to be less than 0.5\,kHz per optical sensor. In this low background environment, IceCube can detect the excess of MeV anti-$\nu_e$ events from a galactic supernova. 

NEMO, an INFN R\&D project in Italy, has been mapping Mediterranean sites and studying novel mechanical structures, data transfer systems as well as low power electronics, with the goal to deploy a next-generation detector similar to IceCube. A concept has been developed with 81 strings spaced by 140\,m. Each consists of 18 bars that are 20\,m long and spaced by 40\,m. A bar holds a pair of photomultipliers at each end, one looking down and one horizontally. As already mentioned, the simulated performance\cite{NEMO} is, not unexpectedly, similar to that of IceCube with a similar total photocathode area as the NEMO concept.

Recently, a wide array of projects have been initiated to detect neutrinos of the highest energies, typically above a threshold of 10 EeV, exploring other experimental signatures: horizontal air showers and acoustic or radio emission from neutrino-induced showers. Some of these experiments, such as the Radio Ice Cerenkov Experiment\cite{frichter} and an acoustic array in the Caribbean\cite{lehtinen}, have taken data; others are under construction, such as the Antarctic Impulsive Transient Antenna\cite{gorham}. The more ambitious EUSO/OWL project aims to detect the fluorescence of high energy cosmic rays and neutrinos from a detector attached to the International Space Stations.

\section*{Acknowledgments}
I thank my IceCube collaborators and Luiz Anchordoqui, Concha Gonzalez-Garcia, Adam Graves, Dan Hooper and Teresa Montaruli for discussions. This research was supported in part by the National Science Foundation under Grant No.~OPP-0236449, in part by the U.S.~Department of Energy under Grant No.~DE-FG02-95ER40896, and in part by the University of Wisconsin Research Committee with funds granted by the Wisconsin Alumni Research Foundation.

\end{document}